\documentclass[aps,twocolumn,superscriptaddress,amsmath,amsfonts]{revtex4-2}

\makeatletter
\renewcommand*{\p@subsection}{\thesection.}
\makeatother

\usepackage{mathtools}
\usepackage{braket}
\usepackage{siunitx}
\usepackage{hyperref}
\usepackage{graphicx}
\usepackage{xcolor}
\usepackage{wrapfig}

\usepackage[framemethod=TikZ]{mdframed}
\newcounter{aside}
\renewcommand{\theaside}{\arabic{aside}}
\newenvironment{aside}[1][]{
    \refstepcounter{aside}
    \begin{mdframed}[
        frametitle={\centerline{Box \theaside:~#1}},
        backgroundcolor=gray!10
    ]
}{
    \end{mdframed}
}

\begin{document}

\title{Mesoscopic ultrafast nonlinear optics---The emergence of multimode quantum non-Gaussian physics}

\newcommand{\SU}{\affiliation{E.\,L.\,Ginzton Laboratory, Stanford University, Stanford, California 94305, USA}}
\newcommand{\NTT}{\affiliation{Physics \& Informatics Laboratories, NTT Research, Inc., Sunnyvale, California 94085, USA}}
\newcommand{\Caltech}{\affiliation{Department of Electrical Engineering, California Institute of Technology, Pasadena, California 91125, USA}}
\newcommand{\MIT}{\affiliation{Research Laboratory of Electronics, MIT, 50 Vassar Street, Cambridge, MA 02139, USA}}
\newcommand{\Cornell}{\affiliation{School of Applied and Engineering Physics, Cornell University, Ithaca, New York 14853, USA}}
\newcommand{\UMassP}{\affiliation{Department of Electrical and Computer Engineering, University of Massachusetts-Amherst, Amherst, Massachusetts 01003, USA}}
\newcommand{\UMassE}{\affiliation{Department of Physics, University of Massachusetts-Amherst, Amherst, Massachusetts 01003, USA}}
\newcommand{\Yale}{\affiliation{Departments of Applied Physics, Yale University, New Haven, Connecticut 06511, USA}}

\author{Ryotatsu Yanagimoto} 
\thanks{These authors contributed equally to this work.\\Email: ryotatsu.yanagimoto@ntt-research.com, edwin.ng@ntt-research.com}
\NTT{}
\SU{}
\Cornell{}

\author{Edwin Ng} 
\thanks{These authors contributed equally to this work.\\Email: ryotatsu.yanagimoto@ntt-research.com, edwin.ng@ntt-research.com}
\NTT{}
\SU{}

\author{Marc Jankowski}
\NTT{}
\SU{}

\author{Rajveer Nehra} 
\Caltech
\UMassP{}
\UMassE{}

\author{Timothy~P.\ McKenna}
\NTT{}
\SU{}

\author{Tatsuhiro Onodera}
\NTT{}
\Cornell{}

\author{Logan G. Wright}
\NTT{}
\Cornell{}
\Yale{}

\author{Ryan Hamerly}
\NTT{}
\MIT{}

\author{Alireza Marandi} 
\Caltech{}

\author{M. M. Fejer} 
\SU{}

\author{Hideo Mabuchi} 
\SU{}

\begin{abstract}
Over the last few decades, nonlinear optics has become significantly more nonlinear, traversing nearly a billionfold improvement in energy efficiency, with ultrafast nonlinear nanophotonics in particular emerging as a frontier for combining both spatial and temporal engineering.
At present, cutting-edge experiments in nonlinear nanophotonics place us just above the \emph{mesoscopic} regime, where a few hundred photons suffice to trigger nonlinear saturation.
In contrast to classical or deep-quantum optics, the mesoscale is characterized by dynamical interactions between mean-field, Gaussian, and non-Gaussian quantum features, all within a close hierarchy of scales.
When combined with the inherent multimode complexity of optical fields, such hybrid quantum-classical dynamics present theoretical, experimental, and engineering challenges to the contemporary framework of quantum optics.
In this review, we highlight the unique physics that emerges in multimode nonlinear optics at the mesoscale and outline key principles for exploiting both classical and quantum features to engineer novel functionalities.
We briefly survey the experimental landscape and draw attention to outstanding technical challenges in materials, dispersion engineering, and device design for accessing mesoscopic operation.
Finally, we speculate on how these capabilities might usher in some new paradigms in quantum photonics, from quantum-augmented information processing to nonclassical-light-driven dynamics and phenomena to all-optical non-Gaussian measurement and sensing.
The physics unlocked at the mesoscale present significant challenges and opportunities in theory and experiment alike, and this review is intended to serve as a guidepost as we begin to navigate this new frontier in ultrafast quantum nonlinear optics.
\end{abstract}

\maketitle

\section{Introduction}

Since the first demonstration of second-harmonic generation (SHG) in 1961~\cite{Franken1961}, nonlinear optics has steadily become ever more nonlinear and energy efficient, making it a ubiquitous driver of progress across modern science and technology, from communications to metrology to sensing~\cite{Diddams2020, Udem2002}.
By one account, such advances can be understood historically through four stages of technological innovation.
The first advance came with the development of phase-matching techniques---and subsequently quasi-phase-matching in particular---which enabled the constructive addition of lightwaves emitted by the nonlinear interaction over long propagation distances~\cite{Maker1962, Giordmaine1962, Fejer1992, Fiore1998}, effectively lifting the micron-scale limitation imposed by phase mismatch.
The second advance came in the form of wave-guiding techniques~\cite{Agrawal2019, Knight1996, Jackel1982}, with which the tradeoff between tight focusing and long interaction can be lifted. 
The third advance was the advent of nonlinear nanophotonics~\cite{Vahala2003, Englund2006, Wang2018, Lu2020, Zhao2022}, where the lithographic definition of subwavelength structures in nonlinear materials enabled tight confinement of light with strong index contrast, increasing the effective intensity and thus the nonlinear efficiency.

\begin{figure*}[tb]
    \centering
    \includegraphics[width=\textwidth]{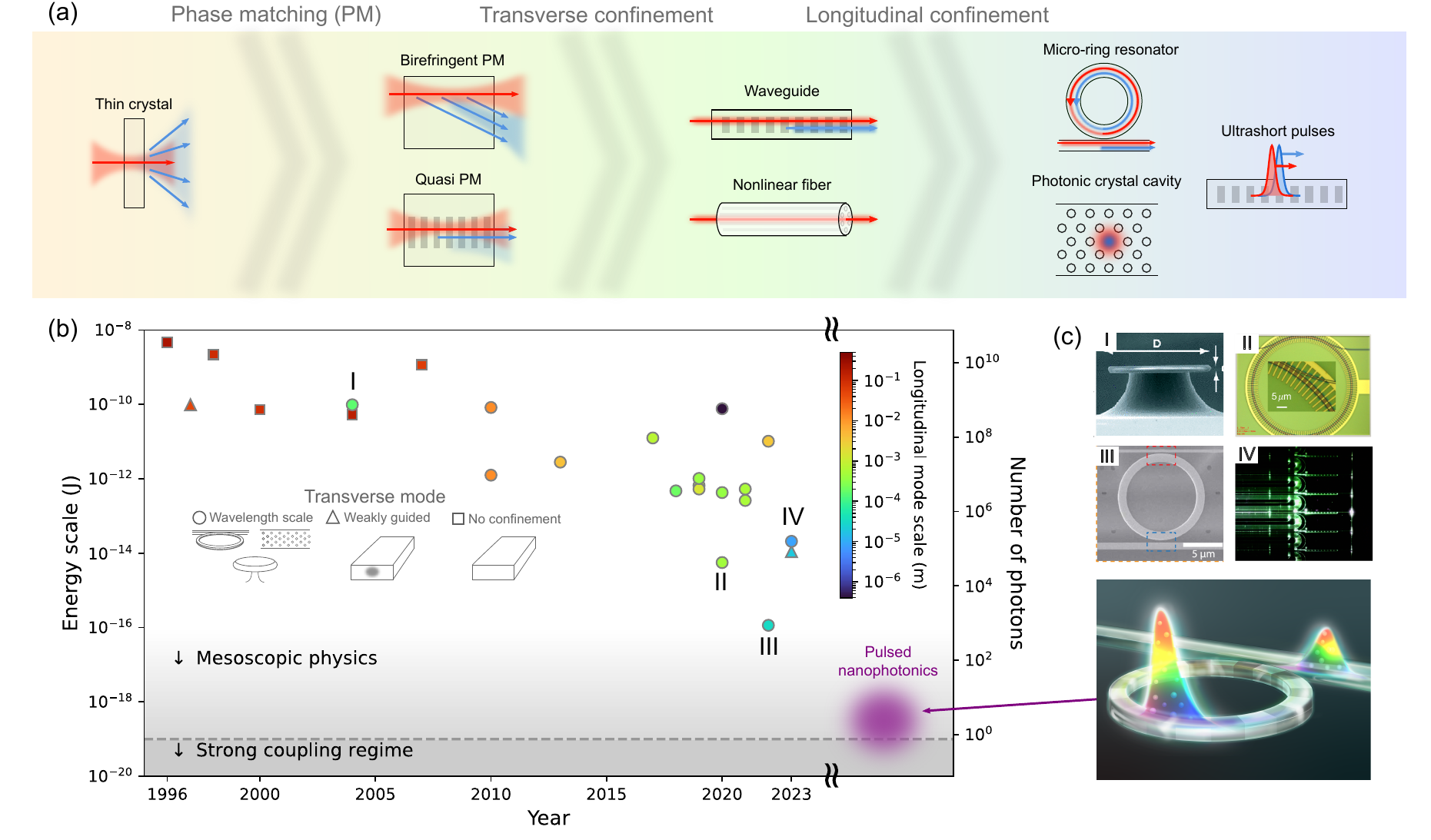}
    \caption{
        Technological progression towards the quantum regime of nonlinear optics.
        (a) Advances in our ability to confine nonlinear-optical interactions in both transverse and longitudinal directions.
        (b) Characteristic energy scales for nonlinear-optical devices.
        For concreteness, we compare the intracavity pump energy $E$ (left axis) or equivalent photon number (assuming \SI{1550}{\nano\meter} light; right axis) at the threshold of optical parametric oscillation; for studies that do not report threshold, we use the formula $E=\hbar\omega_a (\kappa_a/g)^2$, where $g$ is the nonlinear coupling rate, $\omega_a$ is the signal frequency, and $\kappa_a$ is the signal (field) loss rate~\cite{Yanagimoto2022_temporal}.
        Marker shapes and colors depict degrees of confinement in the transverse and longitudinal dimensions, respectively; for pulse-pumped devices, the longitudinal extent is set to the pulse duration.
        Devices considered here include micro-ring waveguide resonators~\cite{Chen2021, Chen2021-photonconversion, Lu2020, Ma2020, Chang2019-GaAs, Chen2019, Bruch2018, Zhao2022, Bruch2019, Ji2017}, whispering-gallery-mode resonators (WGMR)~\cite{Gao2022, Fuerst2010-opo, Fuerst2010-shg, Kippenberg2004}, Fabry-Perot waveguide resonators~\cite{Savanier2013}, free-space resonators~\cite{Samanta2007, Stothard1998, Bosenberg1996, Hatanaka2000, Yu2004}, reverse-proton-exchanged waveguide resonators with quasi-continuous-wave~\cite{Arbore1997} and pulsed~\cite{Jornod2023} pump, and a pulse-pumped nanophotonic resonator~\cite{Gray2023}.
        Selected examples are pictured in (c): in order, (I) a silica WGMR~\cite{Kippenberg2004} (Reprinted figure with permission from Kippenberg et al., \href{https://doi.org/10.1103/PhysRevLett.93.083904}{Phys. Rev. Lett. \textbf{93}, 083904 (2004)}. Copyright 2004 by the American Physical Society.), (II) a periodically poled lithium niobate (PPLN) micro-ring resonator~\cite{Lu2020} (Figure adapted from Lu et al., \href{https://doi.org/10.1364/OPTICA.403931}{Optica \textbf{7}, 12 (2020)}. Copyright 2020 Optica Publishing Group.), (III) an InGaP micro-ring resonator~\cite{Zhao2022} (Figure adapted from Zhao et al., \href{https://doi.org/10.1364/OPTICA.440383}{Optica \textbf{9}, 2 (2022)}. Copyright 2022 Optica Publishing Group.), and (IV) a pulse-pumped PPLN nanophotonic resonator~\cite{Gray2023}.
        Outset of (b) illustrates a device realizing the objective of this review: ultrashort, 3D-confined pulses nonlinearly interacting in the mesoscopic quantum regime.
    } \label{fig:devices}
\end{figure*}

We are now experiencing, in a sense, a fourth development: By combining of all the techniques above with dispersion engineering in nanophotonic waveguides, we can support thousands to millions of frequency modes simultaneously interacting in a single device.
In the time domain, the collective excitation of broadband light manifests as ultrashort (i.e., femtosecond-scale) pulses, producing rich temporal dynamics and further enhancing the nonlinear efficiency via tight confinement of the field in time.
Ultrafast nanophotonics has already opened up new regimes of physical phenomena, device functionalities, and operational paradigms, ranging from micron-scale optical frequency combs~\cite{Kippenberg2018} to ultra-efficient frequency conversion~\cite{Jankowski2022} to broadband quantum light sources~\cite{Kashiwazaki2020, Nehra2022few}.
The result of all the above technological innovations, as exhibited in Fig.~\ref{fig:devices}, is an experimental trend in nonlinear optics over the last several decades featuring a reduction of nearly eight orders of magnitude in the energy scale needed to access optical nonliearities.

Just as remarkably, however, these numbers are also approaching the ultimate physical limit prescribed by quantum mechanics: the energy scale of a single photon, which at \SI{1550}{nm} is on the order of \SI{0.1}{\atto\joule}.
We are therefore confronted with the tantalizing possibility that nonlinear photonics may be on the cusp of unlocking the physics of \emph{strong coupling}, a major milestone widely seen as essential for enabling novel quantum functionalities for applications such as metrology, sensing, and information processing~\cite{Nielsen2000, Degen2017}.
Experimental platforms for ``quantum electrodynamics'' (QED) in which strong coupling has been realized, such as superconducting-circuit QED~\cite{Blais2021}, quantum-dot QED~\cite{Hennessy2007, Press2007}, and atom-cavity QED~\cite{Birnbaum2005, Raimond2001}, are hotly pursued as ``quantum playgrounds'' for testing such functionalities.
However, as the field of quantum engineering advances towards the development of scalable hardware, significant technical challenges on cavity-QED platforms are coming into focus, such as the need for cryogenic cooling and complex setups involving vacuum chambers and many stabilized lasers.
Technologically, it is tempting to ask whether nonlinear nanophotonics might offer a promising alternative or complement to these platforms for quantum science and technology, offering a unique portfolio of room-temperature operability, lithographic scalability, and compatibility with long-distance telecommunications.
Furthermore, just as how complex ultrafast dynamics constitute a fountain of rich physics in classical nonlinear optics~\cite{Herr2014, Grelu2012, Solli2007}, access to many terahertz of bandwidth via advanced dispersion engineering may also enable unique quantum functionalities emerging from the immensely multimode dynamics of interacting photons.

With this motivation, it may appear plausible that we can immediately exploit the full potential of ultrafast quantum nonlinear optics as experimental capabilities advance toward single-photon nonlinearity.
However, it turns out the reality is not quite so simple.
For instance, how does the transition from the macroscopic, classical domain to the microscopic, quantum domain occur?
On the two ends of the spectrum, our physical descriptions of the dynamics take totally different forms: In the classical limit, lightwave propagation is described by classical coupled-wave equations, which are \emph{nonlinear} differential equations, whereas in the microscopic limit, quantum wavefunctions evolve under the Schr\"odinger equation, which is a \emph{linear} differential equation.
These two physical pictures need to be reconciled to understand the intermediate \emph{mesoscopic} regime we inevitably traverse as experimental energy scales continue to target the single-photon level.
Another challenge is the inherently multimode nature of ultrafast nonlinear optics.
For example, to describe a generic quantum pulse with a moderate \num{1000} modes, each populated with at most one photon, quantum mechanics tells us we need a $2^{1000}$-dimensional state space.
In other words, na\"ively speaking, it is a highly nontrivial task even to perform numerical simulations to understand how broadband pulses behave in the quantum domain.
Consequently, many analyses of quantum pulse propagation are based on simplified models drawn from semiclassical intuition, but there are numerous cautionary lessons where failure to account for multimode quantum interactions properly has led to qualitatively wrong predictions~\cite{Shapiro2006, Shapiro2007, Gea-Banacloche2010, Gordon1986}.
To fully leverage and develop the technological opportunities forthcoming in the field, it is thus essential to develop both a deep qualitative understanding of multimode quantum dynamics as well as sophisticated model-reduction capabilities to quantitatively bear it out.

In this review, we explore the transition from classical to quantum ultrafast nonlinear optics and provide one vision for how that transition might experimentally play out in the near future.
In particular, we elucidate the essential physics that characterize nonlinear optics in the mesoscopic regime, where both classical and quantum features in the state coexist and dynamically interact.
The confluence of classical, quantum, and multimode physics suggests a unique opportunity for engineering methodologies from classical optics, quantum optics, and ultrafast optics to overlap and synergize, and we argue that a new program of research should be dedicated to exploring these opportunities.
We discuss some key experimental challenges and capabilities that need further development to access this regime, and we highlight opportunities for altogether new paradigms in quantum engineering, uniquely enabled by the rich phenomenology of mesoscopic quantum optics.

\section{Mesoscopic physics: Navigating the classical to quantum transition} \label{sec:mesoscopic}

\subsection{Nonlinear optics in classical and quantum limits}

Classically, the state of a mode can be represented by its complex amplitude $\alpha$, a single c-number; that is, every amplitude measurement yields $\alpha$ without fail.
Quantum mechanically, however, we know the electric field has no definite value, so two independent amplitude measurements necessarily yield different results, according to some well-defined statistics; that is, there is variance (or perhaps even higher-order statistics) among the measurements.
To handle this, quantum mechanics assigns an operator $\hat a$ to the mode, mathematically defined so that $\alpha=\langle\hat{a}\rangle$, but $\langle\hat{a}^2\rangle$ (closely related to the variance of the amplitude measurements) does not necessarily equal $\alpha^2$.
Thus, $\langle\hat{a}^2\rangle$ is a physical quantity independent from $\alpha$.
Naturally, this argument extends to all higher-order moments too, e.g., $\langle\hat{a}^3\rangle$, $\langle{\hat{a}^{\dagger2} \hat{a}^2}\rangle$, and so on; in general, knowing the full quantum state requires knowing every possible statistic~\cite{Cahill1969}.
Furthermore, in quantum mechanics, the \emph{order} of operators in these moments matter, and noncommutativity leads to uniquely quantum effects, e.g., Heisenberg uncertainty, vacuum fluctuations, and contextuality~\cite{Spekkens2008,Booth2022}.

\begin{figure}[bt]
    \centering
\includegraphics[width=0.5\textwidth]{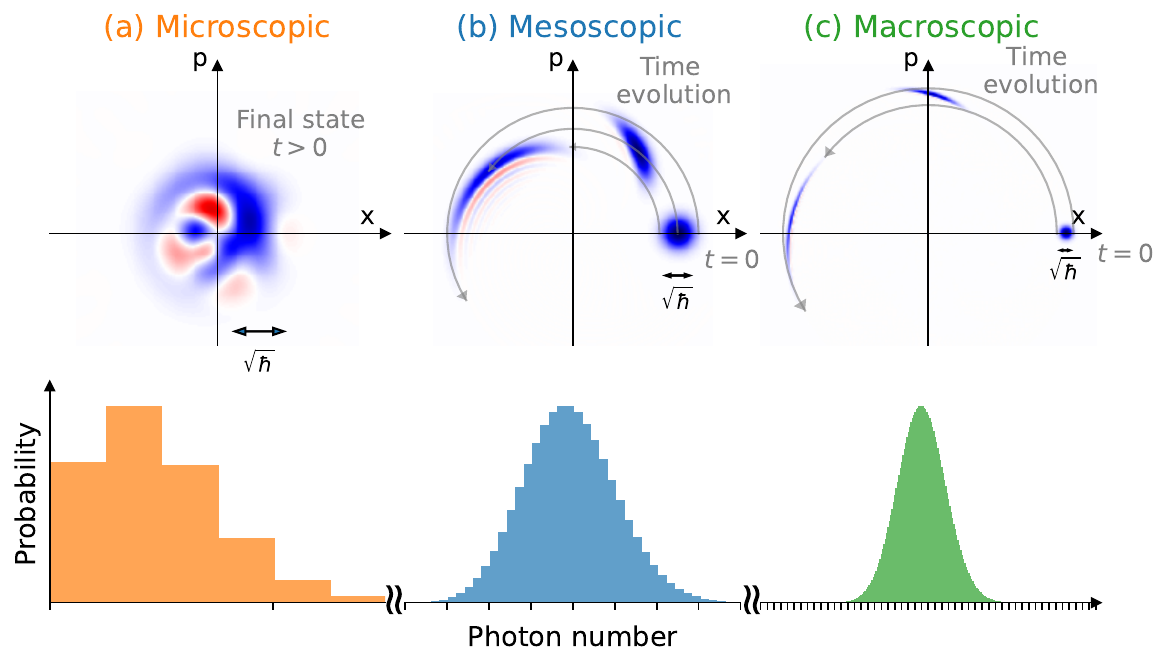}
    \caption{
        Illustration of the phase-space dynamics under Kerr nonlinear interaction for (a) microscopic, (b) mesoscopic, and (c) macroscopic coherent states.
        The bottom graph depicts distribution of photon numbers for each regime.
        The grey arrows represent phase-space flow of corresponding classical mean-field theory, and the black arrows depict the scale of the vacuum fluctuations $\sim\sqrt{\hbar}$.
    } \label{fig:phase-space}
\end{figure}

Obviously, keeping track of all such moments is a daunting task, so when studying deep-quantum phenomena in cavity QED, we often turn to an entirely different representation based on expansions in terms of photon-number states $\ket{n}$.
In this approach, the quantum state is $\ket{\psi} = \sum_{n=0}^\infty c_{n} \ket{n}$,
where we keep track of an infinite sequence $c_n$ instead.
This is an excellent state representation if the physics concerns only a microscopic number of photons since the expansion can be truncated at small $n$.
But the catch is that this confines the state $\ket{\psi}$ to the deep-quantum regime, making it particularly hard to see how the classical world of many photons smoothly transitions to the quantum world of very few: where, e.g., is $\alpha$? For this purpose, it is much more transparent to recall the statistical viewpoint above and treat the quantum state instead as a \emph{distribution} in phase space (a 2D space comprising the real and imaginary parts, or quadratures, of the amplitude).
Just as a probability distribution can be used to calculate all possible statistics for random variables, the phase-space distribution can be used to calculate all possible quantum measurement statistics.

Take, for example, the phase-space portraits, or more technically, \emph{Wigner distributions}~\cite{Wigner1932, Case2008}, shown in Fig.~\ref{fig:phase-space}.
In such plots, the classical first-order moment $\alpha = \langle\hat{a}\rangle$ is simply the centroid of the distribution and is referred to as the \emph{mean-field} amplitude.
But due to quantum uncertainty, there exists a \emph{physical scale} within which the distribution is necessarily blurry: this scale is set by Planck's constant $\hbar$. 
In Fig.~\ref{fig:phase-space}(c), we show a ``classical'' state of the mode, the celebrated coherent state, which consists simply of the centroid $\alpha$ and a symmetric Gaussian distribution around it---the quantum statistics involve only a mean and a variance, from which all other statistics can be computed.
In this depiction, the scale of $\hbar$ is also negligible compared to $\alpha$ in this classical limit, so it makes intuitive sense that we just need to focus on the dynamics of $\alpha$ as the system evolves.

At the other extreme, exotic quantum states such as that shown in Fig.~\ref{fig:phase-space}(a) are characterized by nontrivial features in their Wigner distributions that are comparable or even larger in scale than their mean.
Nonlinear-optical dynamics that generate such features at such scales generally involve strong photon-photon nonlinearities, and this regime involves the proliferation of many nontrivial higher-order moments.
Remarkably, the Wigner distribution in this regime can even feature negative values (subject to certain mathematical constraints), which renders the distribution nonsensical as a conventional probability distribution; in the language of quantum statistics, this property is related to the contextuality of quantum measurements (i.e., the order of operators in the moments)~\cite{Booth2022,Spekkens2008}.
Experimentally, the generation of such features is seen as a holy grail towards procuring physical resources for, e.g., exponential speedups and fault tolerance in quantum computation~\cite{Lloyd1999, Walschaers2021}.

\subsection{A hierarchy of scales in quantum nonlinear optics}
\label{sec:hierarchy-mesoscale}

Intermediate between these two regimes, Figure~\ref{fig:phase-space}(b) depicts how the natural dynamics found in nonlinear optics can take a classical coherent state and \emph{transform} it to a more exotic quantum state.
Looking at the Wigner distribution, the dynamics induce ``flows'' along phase space (in the similar sense as Liouville's equation in classical dynamics).
These flows distort the distribution, which directly translates into changes of the quantum fluctuations and correlations.
Finally, as these distortions become strong enough to generate features on the scale of $\hbar$, the laws of quantum mechanics cause the Wigner distribution to exhibit nonclassical features such as negativity.

\begin{figure}[b]
    \centering
\includegraphics[width=0.5\textwidth]{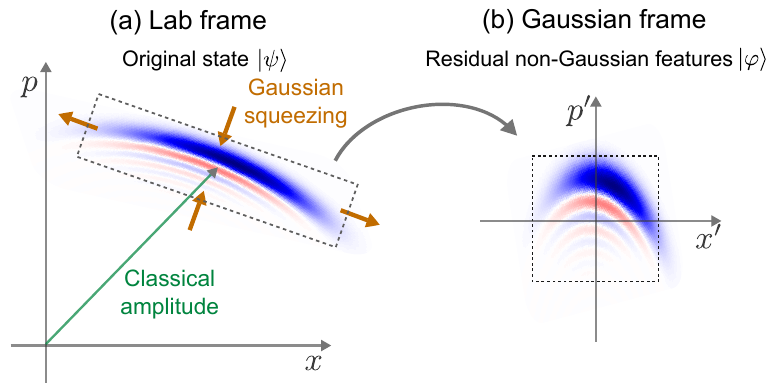}
    \caption{
        (a) In the lab frame, a quantum state generally contains classical features (i.e., mean-field amplitude), semi-classical features (i.e., Gaussian squeezing), and non-Gaussian quantum features.
        (b) Factoring out the dynamics of former two as a Gaussian unitary, non-Gaussian quantum features remain in a Gaussian interaction frame.
    } \label{fig:GIF}
\end{figure}

\onecolumngrid
\vspace{0.2cm}
\begin{aside}[The Gaussian interaction frame] \label{box:GIF}
    Formally, we can, without loss of generality, ``factorize'' the quantum state via a decomposition~\cite{Tezak2017, Yanagimoto2022-non-Gaussian}
    \begin{equation} \label{eq:GIF-ket}
        \ket{\psi} = \hat{D} \hat{S} \ket{\varphi}
    \end{equation}
    with $\hat{D}$, $\hat{S}$, and $\ket{\varphi}$ representing the physics in (i), (ii), and (iii), respectively.
    More precisely, the unitary operators $\hat{D}$ and $\hat{S}$ are the standard displacement and symplectic transformations in quantum optics~\cite{Braunstein2005, Olivares2012}, which manipulate the mean (i.e., $\langle\hat{a}\rangle$) and second-order correlations (i.e., $\langle\hat{a}^2\rangle$ and $\langle\hat{a}^\dagger\hat{a}\rangle$), respectively.
    An insightful way to interpret \eqref{eq:GIF-ket} is to see $\hat{U}=\hat{D}\hat{S}$ as a ``Gaussian'' unitary identifying a ``Gaussian interaction frame''---in this interaction frame, lower-order Gaussian features in the lab-frame state $\ket{\psi}$ are factored out, leaving only nontrivial quantum features in the interaction-frame state $\ket{\varphi}$ (see Fig.~\ref{fig:GIF} for illustration).
    
    A natural choice of $\hat{D}$ and $\hat{S}$ is to take them so that they account for the classical dynamics of the mean-field amplitude and the linearized Gaussian dynamics, respectively.
    More specifically, for a single-mode system, we can introduce parameterizations
    \begin{align}
        \hat{D}^\dagger(\alpha) \hat{a} \hat{D}(\alpha) &= \hat{a} + \alpha,
        &
        \hat{S}^\dagger(\mu,\nu) \hat{a} \hat{S}(\mu,\nu) &= \mu \hat{a} + \nu \hat{a}^\dagger
    \end{align}
    for the annihilation operator $\hat{a}$.
    The evolution of $\alpha$ (equivalently $\hat{D}$) follows the classical coupled-wave equation
    \begin{equation}
        \partial_t\alpha = -\mathrm{i} \partial_{\alpha^*} H(\alpha,\alpha^*),
    \end{equation}
    with $\hat{H}=H(\hat{a},\hat{a}^\dagger)$.
    Then, the evolution of $\mu$ and $\nu$ (equivalently $\hat{S}$) can be obtained by a linearized treatment: We collect first and second order terms of $\hat{D}^\dagger\hat{H}\hat{D}-\mathrm{i}\hat{D}^\dagger\partial_t\hat{D}$ and denote them as $\hat{H}_\mathrm{G}$, with which we get
    \begin{align}
        \partial_t\mu &= \mathrm{i} [\hat{S}^\dagger [\hat{H}_\mathrm{G},\hat{a}] \hat{S}, \hat{a}^\dagger],
        &
        \partial_t\nu &= -\mathrm{i} [\hat{S}^\dagger [\hat{H}_\mathrm{G},\hat{a}] \hat{S}, \hat{a}].
    \end{align}
    Finally, the residual non-Gaussian part $\ket{\varphi}$ evolves in the Gaussian interaction frame under an effective Hamiltonian $\hat{H}_\mathrm{eff}=\hat{U}^\dagger\hat{H}\hat{U}-\mathrm{i}\hat{U}^\dagger\partial_t\hat{U}$; thus the evolution of $\ket\varphi$ can be strongly influenced by the Gaussian unitary $\hat{U}$.
\end{aside}
\twocolumngrid

This generic picture of what happens in phase space can give us significant insight into how classical nonlinear optics transitions to quantum nonlinear optics.
In the former case, the field amplitudes needed to trigger nonlinear effects comprise millions or billions of photons, effectively washing out any observable quantum statistics
As a result, there typically exists a clear hierarchy among the dynamical rates of the following (from faster to slower): (i) mean field affecting mean field (e.g., classical SHG), (ii) mean field affecting quantum fluctuations (e.g., vacuum squeezing), and (iii) quantum fluctuations affecting mean field and quantum fluctuations (e.g., pump depletion solely by amplified vacuum as in parametric generation).
The lowest-order effect (i) is purely classical and corresponds to motion of the centroid of the phase-space distribution, whereas (ii) is the lowest-order nonclassical effect and induces linear squeezings and rotations of the distribution around the mean, as might be captured by a local Jacobian linearizing the flow (see Fig.~\ref{fig:GIF}(a)).
Finally, (iii) induces nonlinear deformations when the flow develops a nontrivial gradient within the extent of the distribution itself, causing most exotic quantum features.
As shown in Fig.~\ref{fig:GIF}(b), such features are most clearly seen as residual non-Gaussian quantum fluctuations after factoring out classical amplitude and Gaussian squeezing.
These visual intuitions and the separation of scales motivate a hierarchical representation of the quantum state itself, as discussed further in Box~\ref{box:GIF}.

\subsection{Non-Gaussianity and the classical-quantum transition}
\label{sec:Gaussian-enhancement}

This hierarchy of scales strongly shapes the ways in which the physics of (iii) can emerge in nonlinear-optical experiments, effectively setting requirements on energy and time scales for generating and observing non-Gaussian quantum features.
With the hierarchy in place, the physics of (iii) can be ignored if the nonlinearity is weak (or equivalently interaction time is short).
In that case, the only states we can generate are \emph{Gaussian states}, so called because their Wigner distributions are Gaussian distributions.
Gaussian states play a central role in modern quantum optics, not only because of their simplicity, but also because they are, to date, the only states that can be deterministically generated in optical experiments with weak nonlinearities.
As such, there is an entire subfield of \emph{Gaussian quantum optics} studying such states as a basic resource~\cite{Braunstein2005, Olivares2012, Asavarant2019, chen2014experimental, Quesada2022}.

To go beyond Gaussian quantum optics, then, we seem to run up against the age-old \emph{strong-coupling problem}: We need larger optical nonlinearities to speed up the physics of (iii) so that non-Gaussianity can arise within timescales faster than decoherence.
At the same time, it is known in classical nonlinear optics that one can accelerate nonlinear dynamics by simply driving the system harder.
For example, we can enhance the power conversion efficiency of SHG by increasing the input pump power.
Could this be a shortcut, unique to nonlinear optics, to generating non-Gaussianity without strong coupling?

The answer is both ``yes'' and ``no.''
Although it is true that states with larger amplitudes experience accelerated development of the exotic quantum features in (iii), the hierarchy of scales implies that the classical and Gaussian features in (i) and (ii) are accelerated also, and even more so.
Specifically, the problem is that dominance of (ii) leads to a strong ``Gaussian background'' that makes the quantum state prone to experimental imperfections, e.g., phase noise and loss.
For instance, in optical parametric generation (OPG), pump depletion can occur even without a signal seed~\cite{Florez2020}, due solely to the brightness of amplified vacuum; this is certainly a manifestation of beyond-Gaussian physics.
However, the strong antisqueezing required to deplete a macroscopic pump field means that the phase-space distribution of the signal is stretched by many orders of magnitude along the amplified quadrature.
The non-Gaussian features appear ``on top'' of this Gaussian one, so they are also macroscopically stretched out in phase space, and they consequently become highly sensitive to loss and noise.
In Sec.~\ref{sec:experiment-tradeoffs}, we specifically explore how this issue quantitatively affects experimental considerations for observing non-Gaussianity.

Thus, in order for non-Gaussian features to emerge, we either need single-photon nonlinearity, or, failing that, \emph{enough} nonlinearity to enhance the development of (iii) with a moderate or \emph{mesoscopic} number of photons.
This second possibility is especially interesting because it involves physics that precisely straddles the classical-to-quantum transition of nonlinear optics, leading to unique phenomenology and functionalities that can emerge at the interface.

\subsection{Quantum optics at the mesoscale}

These considerations motivate us to consider a \emph{mesoscopic} regime of nonlinear optics, where, in the OPG example, only hundreds to thousands of pump photons per pulse suffice to induce pump depletion by amplified quantum fluctuations.
This perspective complements the experimental situation, where rapid advances in nonlinear photonics are pushing closer to this ultra-low energy scale (see Fig.~\ref{fig:devices}).
In this regime, all of the features discussed above---mean-field, Gaussian, and quantum non-Gaussian---\emph{coexist}: They are all significant (none can be neglected), and they dynamically interact within a close hierarchy of scales.

As discussed in Sec.~\ref{sec:hierarchy-mesoscale} (and elaborated in Box~\ref{box:GIF}), there exist ways to represent the quantum state with this hierarchy built-in.
From a modeling perspective, it is precisely in the mesoscopic regime where such a ``multi-scale'' representation is the most warranted, insightful, and efficient.
In this representation, the dynamics within each hierarchical level are driven by the ones above it: Simplifying somewhat, the onset of non-Gaussian physics might look as follows.
First, the mean field, which sits at the highest level, simply follows classical equations of motion.
The mean-field dynamics then drive the growth of quantum fluctuations, whose leading-order dynamics can be represented by the evolution of a Gaussian phase-space distribution.
Finally, the non-Gaussian quantum features evolve under a Schr\"odinger equation (as usual) but driven by nonlinear interactions that directly depend on the dynamics of the previous two hierarchies.
Crucially, the nature of these nonlinear interactions can look dramatically different depending on the exact trajectories taken by the mean-field and Gaussian features.
This process is more formally discussed in Box~\ref{box:GIF}, and we analytically walk through an explicit example this workflow in Box~\ref{box:example-mesoscopic-opa}.

It is worth noting that the physics of the mesoscopic regime is qualitatively distinct from the more conventional cavity-QED-like ``microscopic'' regime, where only a few photons are involved in the dynamics.
In the microscopic regime, even vacuum-level quantum fluctuations are comparable to the mean field, and the hierarchy among mean-field, Gaussian, and non-Gaussian quantum features collapses.
As a result, the ``particle'' nature of photons is more pronounced, motivating a description of the physics via a discrete photon-number basis.
On the other hand, in the mesoscopic regime, the ``oscillator'' aspect of photons remains pronounced, and the state of light is best described via quadrature amplitudes but with non-Gaussian features superimposed, making mesoscopic states a natural carrier of continuous-variable (CV) quantum information~\cite{Braunstein2005} (see Sec.~\ref{sec:photonic-qip}). Moreover, the mesoscopic regime might even be where we can expect to see the greatest quantum complexity:
In the microscopic regime, each mode is only populated by a handful of photons, which makes the Hilbert space dimension grow more slowly than when each mode can accommodate a dozens to hundreds of photons, while in the macroscopic regime, as described above, non-classical features become prone to noise and decoherence.
These considerations suggest the mesoscopic regime is \emph{not} simply a technological compromise (in the NISQ sense~\cite{Preskill2018}), tiding us over until we reach the strong-coupling regime of microscopic cavity-QED-like physics.
Rather, we argue it is an inherently fertile and unique frontier of nonlinear optics that is worthy of investigation in its own right.

\subsection{Classical-quantum co-engineering}
\label{sec:Gaussian-Hamiltonian-engineering}

The presence of strong mean-field and Gaussian features can modify and enhance quantum aspects of nonlinear-optical interactions.
In fact, this idea of \emph{Gaussian Hamiltonian engineering} is well established in quantum photonics where mean-field and Gaussian transformations have been used as powerful tools, e.g., for nonlinear-optical quantum computation~\cite{Nemoto2004, Munro2005, Langford2011, Yanagimoto2020} and the manipulation of quantum decoherence dynamics~\cite{Toyli2016, Rabl2004}.
More recently, innovative quantum-state engineering protocols using composite Gaussian operations, such as the echo conditional displacement protocol~\cite{Eickbusch2022}, have also been demonstrated.
A common template for the design of such protocols is a fixed, initial application of some useful Gaussian transformation \emph{prior} to the nonlinear interaction of interest, which means the Gaussian transformations are largely static and occur in an external, gate-like fashion; this essentially upholds a strong separation of scales between Gaussian and non-Gaussian parts.

In this context, the physics of mesoscopic nonlinear optics may offer an interesting opportunity for \emph{classical-quantum co-engineering}, in which the functional roles played by mean-field, Gaussian, and non-Gaussian physics are less stratified and more coupled.
For instance, non-Gaussian dynamics can often induce residual displacement and squeezing effects, which then re-couple into the dynamics of the classical and Gaussian features as perturbations; in the mesoscopic regime, such perturbations can be of equal magnitude to the non-Gaussian effects, in contrast to the case in Gaussian Hamiltonian engineering where such perturbations are negligibly small.
A concrete example of such a phenomenon is pump depletion in OPG, where the non-Gaussian dynamics induces a reduction in the mean-field displacement of the pump, an effect of the same order as the emergence of non-Gaussian features (see Box~\ref{box:example-mesoscopic-opa} and Ref.~\cite{Yanagimoto2023-cubic}).
Consequently, the dynamics of classical and quantum features can become coupled, potentially translating to new control knobs for engineering devices and quantum states; we elaborate more on such possibilities in Sec.~\ref{sec:qnd}.
It is worth noting that this form of co-engineering among all three levels of the hierarchy is only possible due to the closeness of their respective scales in the mesoscopic regime.
Thus, an equivalent characterization of the mesoscale is the regime in which such classical-quantum co-engineering of photon dynamics can be realized.

\onecolumngrid
\vspace{0.2cm}
\begin{aside}[Mesoscopic optical parametric amplification]
\label{box:example-mesoscopic-opa}
    A central theme in the mesoscopic regime is the interplay between Gaussian and non-Gaussian quantum features, where the former crucially determines how the latter emerges. As a concrete example, we consider a single-mode degenerate $\chi^{(2)}$ interaction with the Hamiltonian
    \begin{align}
    \label{eq:chi2-H}
        \hat{H}=\frac{g}{2}(\hat{a}^{\dagger 2}\hat{b}+\hat{a}^2\hat{b}^\dagger)+\delta\, \hat{a}^\dagger\hat{a},
    \end{align}
    where $g$ is the nonlinear coupling rate, and $\hat{a}$ and $\hat{b}$ are annihilation operators for the signal and pump modes, respectively. The classical dynamics of OPA, described by the coupled equations
    \begin{align}
    &\mathrm{i}\partial_t\alpha=g\alpha^*\beta+\delta\alpha &\mathrm{i}\partial_t\beta=g\alpha^2,
    \end{align}
    are known to change their behavior qualitatively depending on the value of phase-mismatch $\delta$. Does the same hold true in the mesoscopic regime?

    \begin{wrapfigure}{r}{0.5\textwidth}
    \begin{center}
        \vspace{-0.8cm}
\includegraphics[width=0.5\textwidth]{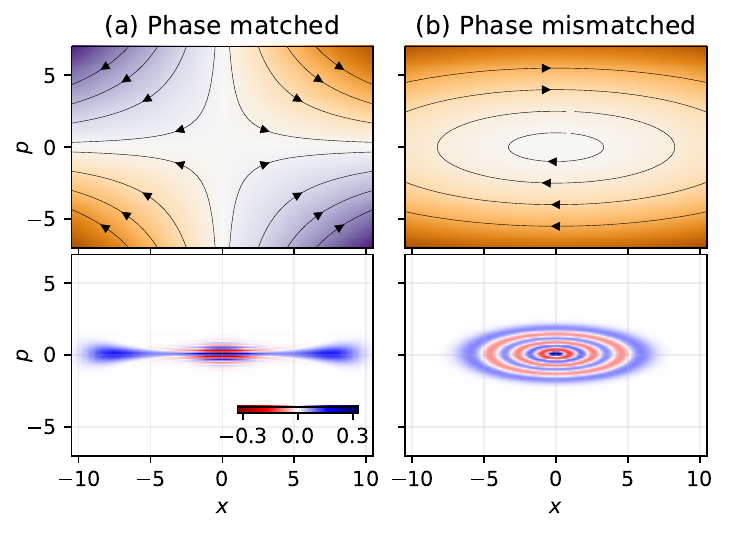}
\vspace{-0.8cm}
\end{center} 
    \caption{
        OPA dynamics in (a) phase-matched and (b) phase-mismatched regimes.
        Upper row: Phase-space flows induced by the Gaussian part of an OPA dynamics.
        Bottom row: Typical signal states produced when conditioned on the quadrature measurement on the pump mode.
    } \label{fig:qnd}
\end{wrapfigure}
    
    When $\delta=0$, OPA is phase matched, and the $x$ quadrature exponentially amplifies while the $p$ quadrature deamplifies. As shown in Fig.~\ref{fig:qnd}(a), such dynamics can be understood as phase-space flows following the equipotential lines of a hyperbolic potential $\propto (x_ap_a+p_ax_a)$, exponentially stretching the phase-space distribution along $x$ and squeezing it along $p$.
    In the limit of large gain, only quantum fluctuations along $x$ are significant, which is, more formally, the statement that $\hat{U}^\dagger\hat{a}\hat{U}=e^{\sqrt{n}gt}\hat{x}_a+e^{-\sqrt{n}gt}\hat{p}_a\approx e^{\sqrt{n}gt}\hat{x}_a$.
    Thus, the interaction-frame Hamiltonian is approximately only a function of $\hat{x}_a$:
    \begin{align}
        \label{eq:phase-matched}
            \hat{H}_\mathrm{eff}\approx g_\mathrm{eff}\hat{x}_a^2\hat{x}_b,
            \quad\text{with}\quad
            g_\mathrm{eff}(t)=ge^{2\sqrt{n}gt}.
    \end{align}
    That is, in this regime of phase-matched OPA, the quantum nature of the pump-signal interaction takes the form of a direct quadrature coupling, which is more insightful than the generic form of \eqref{eq:chi2-H}.
    Additionally, the generation rates of non-Gaussian features by $\hat H_\text{eff}$ are \emph{enhanced} by a dynamical factor $e^{2\sqrt n g t}$. The Hamiltonian \eqref{eq:phase-matched} is known as a cubic quantum nondemolition (QND) interaction~\cite{Budinger2022} and produces a Schr\"odinger's cat state conditioned on quadrature measurement on the pump~\cite{Yanagimoto2023-cubic} (see Fig.~\ref{fig:qnd}(a)).

    On the other hand, for $\delta>g\sqrt{n}$, OPA is phase-mismatched, and the parametric gain periodically oscillates without exponential growth.
    In this regime, the Gaussian part of the phase-space dynamics corresponds to periodic orbits along stretched ellipses as shown in Fig.~\ref{fig:qnd}(b).
    Mathematically, this can be described as a detuning generated by squeezed operators, i.e., $\hat U = \exp(-\mathrm{i} \Delta \hat A^\dagger \hat At)$, where $\hat{A} = \cosh u \, \hat{a} - \sinh u \, \hat{a}^\dagger$, with $u=\frac{1}{2}\tanh^{-1}(g\sqrt{n}/\delta)$ and $\Delta=\sqrt{\delta^2-g^2n}$.
    Since the rotational rate is much faster than the non-Gaussian quantum dynamics in the mesoscopic regime, all the phase-sensitive terms in the Hamiltonian average out, which effectively projects the non-Gaussian dynamics into the eigenspace of $\Delta\hat{A}^\dagger\hat{A}$, i.e., squeezed photon-number states.
    As a result, the interaction-frame Hamiltonian becomes~\cite{Qin2022,Yanagimoto2023-qnd}
    \begin{align}
        \label{eq:phase-mismatched}
            \hat{H}_\mathrm{eff}\approx g_\mathrm{eff}\hat{A}^{\dagger}\hat{A}\hat{x}_b,
            \quad\text{with}\quad
            g_\mathrm{eff}=g\sinh(2u),
    \end{align}
    where we have ignored terms that can be eliminated by additional trivial Gaussian transformations. The form of \eqref{eq:phase-mismatched} is known as ponderomotive coupling, which enables, e.g., photon-number-resolving QND measurement. Conditioned on the pump quadrature value, the signal is projected to a squeezed Fock state~\cite{Yanagimoto2023-qnd} (see Fig.~\ref{fig:qnd}(b)).

\end{aside}
\twocolumngrid

\section{Multimode physics}
\label{sec:multimode}
Nonlinear optics is inherently multimode.
Physically this is because, to a first approximation, the nonlinear susceptibility describes a spatially \emph{local} coupling between fields, whereas the normal modes of many optical systems are usually \emph{nonlocal}, for example the frequency modes of a resonator.
Consequently, optical nonlinearity tends to scatter photons among many such normal modes, and special engineering effort is required to obtain effective single-mode nonlinear interactions.

This intrinsic multimodedness is both a blessing and a curse:
On one hand, multimode interactions complicate the dynamics, resulting in significant deviations from the single-mode theory, e.g., as observed in the efficiency of nonlinear frequency conversion~\cite{McKenna2022}.
On the other hand, access to a massive number of modes is central to some of the most unique physical mechanisms in the physics of light, such as modulation instability~\cite{Solli2007}, mode locking~\cite{Grelu2012}, and supercontinuum generation~\cite{Dudley2009}, and it is also the key to photonic technologies like highly multiplexed communication links, broadband spectroscopy~\cite{Coddington2016}, and frequency metrology~\cite{Takamoto2005}.
Furthermore, as our abilities to control and probe multimode systems improve, there has been a resurgence of interest in their quantum properties, as evidenced by recent interesting experiments studying multimode Gaussian physics on platforms amenable to linear analysis of quantum correlations, such as soliton microcombs~\cite{Bao2021,Guidry2022,Guidry2023}, supercontinuum generation~\cite{Uddin2023}, and pulsed squeezing~\cite{Triginer2020}.
In the following subsections, we survey the landscape ahead and discuss multimode nonlinear optics in the context of mesoscopic non-Gaussian physics, drawing attention to some recent developments and open questions.

\subsection{Ultrashort pulses can enable strong coupling}
\label{sec:ultrashort-pulse}

As noted before in this article, the realization of strong nonlinearites is an outstanding challenge in optics.
To enhance nonlinear coupling, it is essential to confine photons to a small volume, e.g., via whispering-gallery disk resonators, micro-ring waveguide resonators, or photonic-crystal cavities.
However, in most photonic platforms, there is an important empirical tradeoff between the mode volume (and thus the nonlinear rate $g$) and the loss rate $\kappa$, in effect limiting the overall strong-coupling figure of merit $g/\kappa$.
For instance, decreasing the size of a ring resonator increases bending loss and surface roughness loss. Consequently, even the best nonlinear coupling realized to date remains two orders of magnitude from the strong-coupling threshold $g/\kappa \sim 1$.

Ultrashort pulses offer an interesting way around this tradeoff.
Intuitively, it is natural to see an optical pulse as a ``flying cavity'' whose size is set by the pulse duration.
To the extent this is true---i.e., up to caveats next discussed in Sec.~\ref{sec:quantum-multimode}---an ultrashort pulse can therefore realize tight temporal (longitudinal) field confinement without shrinking the size of the physical resonator, circumventing the loss-nonlinearity tradeoff.
Combined with transverse field confinement enabled by nanophotonic waveguides, reaching the strong-coupling threshold $g/\kappa\sim 1$ appears plausible in principle, based on the state of the art in thin-film lithium niobate (TFLN) (see Sec.~\ref{sec:tfln}); with more futuristic parameters, $g/\kappa\sim 100$ may even be possible~\cite{Yanagimoto2022_temporal}.

\subsection{Quantum multimode effects are more subtle than classical}
\label{sec:quantum-multimode}

While the use of multimode dynamics is standard in classical nonlinear optics, quantum treatments of nonlinear optics have historically been strongly influenced by cavity QED and are therefore often formulated as single-mode models.
Such single-mode quantum models can provide useful theoretical intuition, but special care is needed when extrapolating that intuition to real physical systems, particularly regarding when and how multimode effects can emerge.
For an $M$-mode system, the classical mean field consists only of $\mathcal{O}(M)$ parameters, while $n$th-order correlations of the quantum fluctuations require $\mathcal{O}(M^n)$, suggesting that in the quantum regime, there are simply more ways in which multimode effects can participate in the dynamics.

For example, consider vacuum squeezing in a waveguide, pumped by continuous-wave light.
While the mean-field dynamics seem completely static and single-mode, the second-order quantum correlations among signal photons exhibit highly multimode squeezing with broadband correlations~\cite{Quesada2022}.
In other words, we \emph{cannot} use a single-mode Hamiltonian $\hat{H}_\mathrm{sqz}\propto(\hat{a}^{\dagger2}+\hat{a}^2)$ to describe travelling-wave squeezing.
In fact, enforcing single-modedness in squeezed light generation is known to be a nontrivial and important task in quantum information science, which requires appropriate pump pulse shaping, dispersion engineering, poling apodization, and gain control~\cite{Grice1997, Ansari2018, Houde2023, Gorbach2023}.

There exist many other cautionary tales where failure to account for multimode quantum effects can lead to qualitatively incorrect predictions.
An illuminating case is provided by Shapiro in Ref.~\cite{Shapiro2006} under the provocative title ``Single-photon Kerr nonlinearities do not help quantum computation''.
Preceding this work, it was proposed that two-qubit entangling gates might be realized in nonlinear optics using a Kerr cross-phase modulation (XPM) Hamiltonian $\hat{H}_\mathrm{XPM}\propto\hat{a}^\dagger\hat{a}\hat{b}^\dagger\hat{b}$~\cite{Chuang1995}.
However, Shapiro argued that the XPM interaction as proposed is inherently multimode, and hence the na\"ive single-mode Hamiltonian $\hat{H}_\mathrm{XPM}$ is not a sufficient description.
Using the proper broadband waveguide Hamiltonian, Shapiro showed that multimode interactions act as effective decoherence channels, preventing the realization of high-fidelity quantum gates in that setup, no matter the strength of the nonlinearity.
Rather, as we further discuss below, more elaborate techniques are required to confine the nonlinear interaction to a single-mode subspace.

With these observations, we see that the analogy made in Sec.~\ref{sec:ultrashort-pulse} between a ultrashort pulse and a ``flying cavity'' is not entirely accurate, being based on a single-mode picture.
Rather, we must ask how single-mode physics might be recovered, and to what extent.
For instance, the quantum dynamics of an optical soliton can exhibit approximately single-mode dynamics due to the balance between multimode nonlinear interactions and linear dispersion, but at the same time, this is only true up to quantum soliton evaporation~\cite{Lai1989_b, Wright1991, Korolkova2001}.
Interactions between wavepackets with large group-velocity mismatch can also be approximately single-mode and can enable high-fidelity quantum-gate operations~\cite{Xia2016, Viswanathan2018, Babushkin2022}.
In the mesoscopic regime, non-Gaussian quantum features tend to first emerge along supermodes with strong levels of squeezing, suggesting that non-Gaussianity could, at least initially, be contained within a small subspace spanned by the principal squeezing supermodes~\cite{Yanagimoto2022-non-Gaussian}.
Finally, truly single-mode pulse dynamics can be enforced in a state-independent way using a \emph{temporal trap}, which consists of an effective potential in the temporal domain formed by the nonlinear phase shift imposed by an auxiliary trapping pulse~\cite{Yanagimoto2022_temporal}.
We emphasize that it is in the context of this engineered single-mode dynamics that the arguments in Sec.~\ref{sec:ultrashort-pulse} regarding temporal confinement and strong coupling become valid.

\subsection{Model reduction for multimode non-Gaussian dynamics}
\label{sec:model-reduction}

Numerical simulations play an indispensible role in elucidating the essential behavior of complicated physical systems, but applying numerical methods to study the quantum dynamics of photons in the multimode and non-Gaussian regime can be quite challenging.
Na\"ively, the quantum state space of $M$ modes with at most $N$ photons in each mode is $\mathcal O(N^M)$-dimensional.
It is therefore clear that we need sophisticated \emph{model reduction} techniques to obtain efficient numerical representations of quantum states and means of computing their dynamics.

At a high level, the complexity of multimode quantum systems can be broken down into three key elements: the number of modes, the degree of non-Gaussianity, and the degree of entanglement.
When all three elements are present at significant levels (e.g., within fault-tolerant quantum computers or simulators at scale), quantum dynamics can be exponentially complicated and intractable for conventional computers.
However, practical limitations or operational choices often place limits on one or more of these elements, providing an opportunity to develop tractable reduced models.
In this context, multimode mesoscopic systems---in which non-Gaussianity is naturally limited to an intermediate level---is a prime candidate for model reduction methods.
Below, we review some numerical techniques that have proven useful for initial forays into the multimode non-Gaussian dynamics of mesoscopic nonlinear-optical systems.

\paragraph{Phase-space trajectory methods}
In these methods, the quantum state is given an alternative representation as a (possibly quasi-probability) distribution in phase space.
Then, the quantum dynamics of the phase-space distributions can be reconstructed using Monte-Carlo methods, e.g., in the form of stochastic differential equations (SDEs).
Among the possible choices of phase-space methods, the positive-$P$ method~\cite{Drummond1980_p} and truncated-Wigner method~\cite{Sinatra2002} are particularly well known.
Technically, the truncated-Wigner method cannot capture negativities in phase space, whereas the positive-$P$ method could.
Both methods become accurate under sufficiently large dissipation, which makes them ideal for studying classical-quantum transitions in driven-dissipative systems~\cite{Propp2023,Deuar2021}.
Note that the corresponding SDEs can become unstable and chaotic when the states become strongly non-Gaussian, leading to slower numerical convergence.

\paragraph{Truncated cumulant expansion}
For a Gaussian quantum state, all but first- and second-order cumulants vanish, suggesting that slightly non-Gaussian states may be well captured by taking a cumulant expansion and applying a finite \emph{truncation} at some order.
For an $M$-mode system and an $n$th order cumulant expansion, the state-space complexity of the model becomes $\mathcal{O}(M^n)$.
Such cumulant expansion has been successfully applied to few-mode cavity-QED and nonlinear-optical systems~\cite{Schack1990, Huang2022, Huang2023}.
However, even at fixed $n$, the relatively expensive polynomial scaling of $\mathcal{O}(M^n)$ make such methods challenging for systems with a large number of modes $M$.
At the same time, as discussed further in Sec.~\ref{sec:augumented}, the dynamics of such models also suggest an interesting interpretation of quantum effects in terms of a ``quantum-augmented configuration space''.

\paragraph{Non-Gaussian supermode model}
In this model-reduction technique specifically developed for studying mesoscopic physics, we first approximately obtain the Gaussian dynamics of the system.
We then make use of the hierarchy of dynamical scales discussed in Sec.~\ref{sec:hierarchy-mesoscale} to pick out a small set of modes that have strong Gaussian excitations, which we understand to be the most likely modes in which non-Gaussian physics would arise.
We can then devote computational resources to performing quantum simulations of these ``principal squeezing supermodes'', while approximating all other modes as being in Gaussian states.
A full quantum simulation of mesoscopic broadband OPG has been demonstrated using this approach~\cite{Yanagimoto2022-non-Gaussian}.

\paragraph{Matrix product states}
Unlike the other methods above, which rely on limited non-Gaussianity, this approach aims at leveraging limited entanglement.
It is known that in one-dimensional quantum many-body systems, the amount of entanglement generated is limited, and an alternative representation of the quantum state in the form of a matrix-product state (MPS) provides an efficient description of that entanglement~\cite{Vidal2002, Vidal2003}.
While the MPS framework is intensively used in the field of many-body physics, the one-dimensional nature of waveguide propagation makes it also amenable to MPS simulation, and quantum nonlinear optics in the few-photon (microscopic) regime can indeed by simulated with such methods~\cite{Yanagimoto2021_mps}.
However, note that the presence of strong squeezing in multimode dynamics incurs a high (though polynomial) numerical cost, making MPS methods less suitable for mesoscopic physics with stronger Gaussian features.
It is an interesting open question whether one can combine MPS methods with, say, the Gaussian interaction frame to obtain efficient representations of just the non-Gaussian features.

While the techniques mentioned above have pushed the boundaries of our understanding beyond conventional Gaussian quantum optics, they are far from enough to fully cover all the possible states of mesoscopic non-Gaussian light that might be realized in future experiments.
For instance, how might we model mesoscopic supercontinuum generation where millions of modes undergo complicated non-Gaussian quantum dynamics?
Clearly, further research in sophisticated model reduction is acutely needed to even map out the landscape of possibilities in mesoscopic nonlinear optics. 

\section{What does this all mean for experiments?}
\label{sec:experiment}

Our discussion thus far has focused on the generic features of multimode mesoscopic physics in order to emphasize the fact that such behavior can, in principle, arise in all nonlinear-optical systems.
However, to actually observe and harness these physics, a given experimental setup must be able to access the energy scales defining the mesoscopic regime, support enough modes to produce interesting complexity, and allow flexible access to and control of its dynamics, all of which depend critically on device design and hardware architecture.
In this section, we survey the experimental factors most relevant to achieving these prerequisites, highlighting the main achievements that have taken us to the cusp of mesoscale nonlinear optics, some known experimental challenges (particularly in nanofabrication and dispersion engineering), and the need for new techniques at the intersection of ultrafast, quantum, and nonlinear optics.

We first translate the physical boundaries of the mesoscale regime into concrete figures of merit for nonlinearity and loss that can be applied to experimental devices; unsurprisingly in light of the discussion thus far, mesoscopic physics arises in a ``Goldilocks zone'' in pump energy and propagation time.
Second, we review some of the key challenges and successes in the nanofabrication of nonlinear optical devices: This is a notoriously technical subject,
but we aim to give a glimpse by highlighting some established milestone results---from low-loss waveguides to periodic poling to wafer-scale fabrication---in thin-film lithium niobate, a major candidate material platform for quantum nanophotonics.
In a similar spirit, we also review the main challenges in dispersion engineering of nanophotonic waveguides and resonators, which enables not only the rich physics of mesoscopic broadband fields but also entirely new regimes of ultrawide-bandwith operation in quantum and coherent information processing, as discussed later in Sec.~\ref{sec:photonic-qip}.
Finally, we conclude with a discussion on what we see as an interesting opportunity in experimental photonics: the development of new techniques and capabilities that can simultaneously manipulate classical and quantum features in ultrafast pulses in the mesoscopic regime, combining the spectral-temporal resolution of ultrafast optics with the ultralow-power operability and coherence of quantum optical operations.
Taken together, we believe these experimental challenges and opportunities represent a burgeoning subfield of photonics that draws from the disciplines of analog computation, quantum information, nonlinear optics, ultrafast optics, and quantum optics.

\subsection{Experimental figures of merits and tradeoffs}
\label{sec:experiment-tradeoffs}

As discussed in Sec.~\ref{sec:mesoscopic}, quantum features arise by nonlinear interactions among photons, and the degree of interaction is governed by the product between a nonlinear rate $g$ and an interaction time $t$.
For non-Gaussian quantum features to form, the quantity $\tau = gt$ needs to reach a certain threshold value $\tau_\mathrm{th}$.
This means strong nonlinearity is \emph{not} the whole story, and experimental considerations need to take both nonlinearity and interaction time into account.
For instance, some emerging two-dimensional systems feature ``giant'' optical nonlinearity $g$ but only offer vanishingly small interaction time $t$, leading to insufficient $\tau$~\cite{Khurgin2023}.
In single-pass devices (e.g., a waveguide), the interaction time is set by the time-of-flight and hence device length.
In this context, \emph{resonators} can provide significant leverage, letting us recycle the nonlinear region and obtain much longer interaction lengths for the same device footprint.
A related technique is the use of ``slow light'' via  electromagnetically induced transparency, where resonant atomic transitions extend the photon-photon interaction time~\cite{Cantu2020}.

Crucially, for schemes that rely on increasing $t$, photon loss almost invariably becomes the limiting process.
In order to access non-Gaussian quantum features, the photon loss rate $\kappa$ needs to be small compared to $g$ so that nonlinear dynamics occur faster than decoherence.
The resulting figure of merit $g/\kappa$ suggests that, again, strong nonlinearity is not the whole story, and what matters is its ratio with the loss.
For instance, epsilon-near-zero materials can offer extremely large nonlinearity~\cite{Alam2016}, but their large losses have hindered applications to quantum optics.
In nanophotonics, decreasing resonator volume can enhance nonlinear coupling, but doing so also usually increases losses due to, e.g. bend radius and surface roughness~\cite{Zhao2022}.
As discussed in Sec.~\ref{sec:ultrashort-pulse}, the use of ultrashort pulses may offer a means to circumvent this particular nonlinearity-loss tradeoff. The ability to dynamically couple light in and out of a resonator, e.g, via nonlinear-optical switching~\cite{Heuck2020b}, can be a powerful technique to extend $t$ without compromising the operation bandwidth.

Finally, we note that there are additional considerations, unique to the mesoscopic regime of operation, that modify the above requirements on $\tau_\mathrm{th}$ and $g/\kappa$ in ways that can be highly dependent on the dynamics at hand.
Due to the Gaussian enhancement of nonlinear dynamics as discussed in Sec.~\ref{sec:Gaussian-enhancement}, we can generally reduce $\tau_\mathrm{th}$ by stronger pumping (i.e., using more photons), which can alleviate the device-length requirement for single-pass devices. At the same time, large Gaussian features also enhance decoherence, and the net requirement for $g/\kappa$ to retain coherent non-Gaussianity is determined by the dynamical competition between the Gaussian enhancements of nonlinearity and loss.
As a result, these experimental requirements could vary by orders of magnitudes depending on \emph{how we operate them}, ranging from a similar requirement to strong coupling $g/\kappa \sim 1$ to even down to $g/\kappa \sim \num{e-4}$~\cite{Yanagimoto2020}. In Box~\ref{box:experimental_opa}, we illustrate these important but subtle considerations for the example of mesoscopic OPA.

\onecolumngrid
\vspace{0.2cm}
\begin{aside}[Experimental considerations for mesoscopic OPA]
\label{box:experimental_opa}
    To elaborate on how mesoscopic physics can affect the experimental figures of merit for observing non-Gaussian quantum effects, let us consider again the example of a cubic QND interaction in phase-matched OPA.
    In a typical setup for this example, we are interested in resolving the conditional pump displacement $\zeta$ in the interaction $\exp(-\mathrm{i}\zeta\hat{x}_a^2\hat{x}_b)$.
    Suppose we want to achieve a nominal value of $\zeta = 1$ (e.g., for reasonable single-shot resolution under coherent-state statistics for the pump).
    Considering the Gaussian-enhanced interaction rate in \eqref{eq:phase-matched}, we find that the threshold time is $\tau_\mathrm{th}=gt_\mathrm{th}\approx \frac{1}{2\sqrt{n}}\log(2\sqrt{n})$.

\begin{wrapfigure}{r}{0.5\textwidth}
    \begin{center}
        \vspace{-0.8cm}
\includegraphics[width=0.5\textwidth]{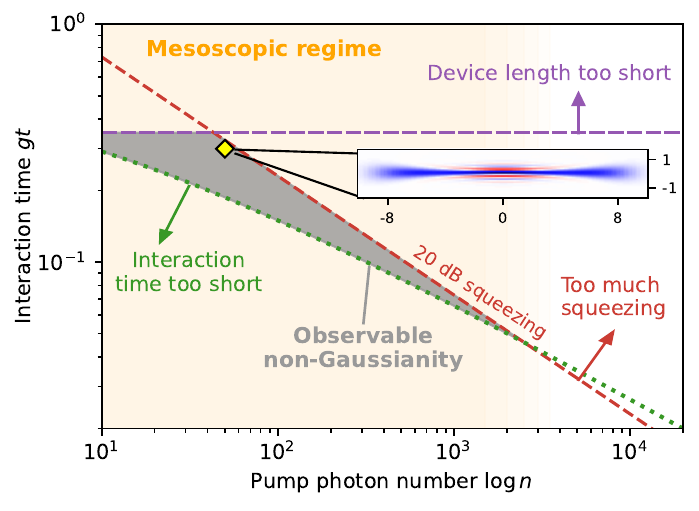}
\vspace{-0.8cm}
\end{center} 
    \caption{
        The grey shaded region indicates the parameter regime in which robust non-Gaussian quantum features could in principle be observed experimentally in mesoscopic OPA, provided low enough loss rates. Red dotted line: Upper tolerance limit of parametric gain $G_\mathrm{max} = 100$, corresponding to $20$ dB of squeezing. Green line: Threshold (minimum) interaction time $g t_\mathrm{th}$ needed to develop non-Gaussianity. Purple line: Total interaction time set by the effective device length, due to hardware limits or the effective decoherence rate.
        Inset: The Wigner function of a typical signal state obtained after homodyne on the pump, for parameters indicated by the yellow diamond.}
        \label{fig:non-Gaussian}
    \end{wrapfigure}
    
    As mentioned in the main text, however, this is not the only rate that is enhanced.
    We observe that (anti)squeezing is the dominant Gaussian effect in this example (there is no displacement for vacuum signal input).
    The energy contained in this Gaussian feature is quantified by the power gain $G$ of the OPA, which is reached after an interaction time $gt_G\approx\frac{1}{2\sqrt{n}}\log G$.
    Above some level of gain $G_\mathrm{max}$, there is so much antisqueezing that non-Gaussian quantum features cannot survive experimentally reasonable levels of noise and loss.
    For example, phase sensitivity of quadrature-squeezed states exhibit Heisenberg scaling~\cite{Qin2023, Nielsen2023}, making the quantum features extremely prone to phase noise.
    This condition can be formulated as $t_\mathrm{th}\leq t_{G_\mathrm{max}}$, which implies the pump energy (used to enhance the nonlinearity) should not exceed $n\leq G^{2}_\mathrm{max}/4$.
    For instance, with $G_\mathrm{max}=100$, corresponding to $\SI{20}{dB}$ of gain or (anti)squeezing, we can tolerate $n\leq 2500$, which marks the upper boundary of the mesoscopic regime as depicted in Fig.~\ref{fig:non-Gaussian} for this example.
    These considerations carve out a ``Goldilocks zone'' in Fig.~\ref{fig:non-Gaussian}, in which non-Gaussian quantum features could in principle be observed robustly.
    
    All of this, however, is assuming that the photon losses are sufficiently small, i.e., that the characteristic decoherence time $t_\mathrm{loss}$ is longer than the threshold propagation time $t_\mathrm{th}$.
    Here, it is essential to note that the decoherence rate of a quantum state increases proportionally to the size of quantum fluctuations, and thus, states in the mesoscopic regime can actually decohere faster than microscopic few-photon quantum states. Mathematically, one can also view this as a Gaussian enhancement, not of the nonlinear dynamics, but of the loss itself: Starting from a native single-photon loss rate $\kappa$, the decoherence time for the cubic QND example is enhanced to $gt_\mathrm{loss}\approx\frac{1}{2\sqrt{n}}\log(2\sqrt{n}g/\kappa)$, and the requirement for loss can be formulated as $t_\mathrm{loss}>t_\mathrm{th}$. 
    Yet in spite of the enhanced loss, this condition is no worse than the usual strong-coupling condition $g/\kappa \ge 1$, as the enhanced loss is precisely balanced by the enhanced nonlinearity. In fact, the balance can even be made to tip in our favor: simply adding simply adding some pump squeezing can relax this to $g/\kappa\geq 0.1$~\cite{Yanagimoto2023-cubic}, underscoring the subtlety of ``nonlinear enhancements'' at the mesoscale.
  
\end{aside}
\vspace{0.2cm}
\twocolumngrid

\subsection{Materials and nanofabrication: Perspectives from thin-film lithium niobate} \label{sec:tfln}

Many promising platforms exist for quantum and ultrafast nanophotonics, all taking advantage of the miniaturization of photonic structures in materials with strong optical nonlinearity, with some of the most established platforms being silicon-based photonics~\cite{Ji2017,Guidry2022,Guidry2023,Moss2013} and III-V semiconductor systems~\cite{Zhao2022,Chang2019-GaAs,Baboux2023}. Alongside these major players, thin-film lithium niobate (TFLN) stands as a major emerging candidate material platform~\cite{Zhu2021} due to its strong $\chi^{(2)}$ and $\chi^{(3)}$ nonlinearities, the electro-optic effect, wide materials transparency range, large index contrast for compact waveguides, and experimental demonstrations of low loss waveguides. Many platforms exist for nonlinear nanophotonics, but few materials encompass all of these features natively and monolithically.

The material loss limit of TFLN wafers prepared with the ion slicing technique has been measured to be around $\SI{0.2}{dB/m}$ in the telecommunications band, which corresponds to a resonator quality factor of 160 million~\cite{Sham-Ansari2022}. Typical demonstrations of strongly-guided ring resonators feature quality factors between 1 million and 5 million~\cite{ShamsAnsari2022, Li2023}, with outliers reporting greater than 10 million~\cite{Zhang2017-highQ}. An exception to this trend is TFLN weakly-guided whispering gallery resonators which are fabricated by chemical-mechanical polishing techniques and can have quality factor greater than 100 million~\cite{Gao2022}. However, these devices do not readily allow for monolithic integration compared to wavelength-scale nanophotonic waveguides which are necessary for ultrafast dispersion engineering.

Lithium niobate allows for periodic poling to provide quasi-phase matched $\chi^{(2)}$ interactions with mm to cm interaction lengths. The grating is formed by periodically inverting the orientation of the ferroelectric polarization of the crystal, patterning the sign of the nonlinear susceptibility tensor, enabling `quasi-phase-matched' interactions that compensate for the momentum mismatch between nonlinearly interacting waves (e.g., fundamental and second-harmonic)~\cite{Hum2007}. This is a major advantage compared to other material platforms that either entirely lack support for quasi-phase matching or rely upon orientation patterning~\cite{Eyres2001} which is less developed for nanoscale waveguides. Gratings in TFLN with sub-micron periodicity have been demonstrated~\cite{Nagy2020}, and many works with $\sim$ $\SI{4}{\micro\meter}$ periods have been demonstrated for supporting telecom wavelengths~\cite{Wang2018}. It now appears that periodic poling has the resolution to support quasi-phase matched nonlinear optics from visible to the mid-infrared wavelengths.

Many of the most important functionalities have been demonstrated in the TFLN platform with individual devices that are fabricated with low-yield processes. Current efforts seek to expand fabrication methods to scalable and repeatable wafer-scale fabrication techniques. Major challenges of wafer-scale fabrication are the uniformity of the lithium niobate film thickness across the wafer and uniformity of the etch processes. Carefully dispersion-engineered nonlinear circuits are particularly sensitive to such fabrication nonuniformity but process improvements are within reach to produce quantum nonlinear optical circuits at scale with 100 to $\SI{200}{mm}$ diameter wafers.

\subsection{Challenges and opportunities in dispersion engineering}

For single-frequency devices in a given hardware platform, the main experimental challenges are as described in Sec.~\ref{sec:experiment-tradeoffs}: Loss rates must be decreased and nonlinear coupling rates must be increased.
But as discussed in Sec.~\ref{sec:multimode}, the use of ultrafast (i.e., broadband) pulses can offer many new advantages, from conceptual ones like the emergence of complex multimode quantum physics to practical ones like mitigating tradeoffs between loss and mode volume in resonators.
Such opportunities come with new challenges.
The broadband nature of ultrafast nonlinear optics means that the dispersion relations of the nonlinear waveguide now play a dominant role in the operation of the device; indeed, as mentioned in Sec.~\ref{sec:multimode}, inadequate understanding or control of the multimode physics can open up deleterious decoherence channels~\cite{onodera2022nonlinear} or disrupt quantum effects that are otherwise expected to occur in a single-mode model~\cite{Shapiro2006}.

Early work in ultrafast quantum optics established the formal relationship between dispersion relations and the physics of unsaturated (i.e., linearized) multimode dynamics, namely, the discovery and characterization of pulsed (or temporal) eigenmodes.
These efforts led to a multimode, Gaussian quantum theory for unsaturated processes such as SPDC, undepleted OPA, and below-threshold OPOs pumped by ultrafast pulses, allowing us to decompose their dynamics into a fixed basis of temporal waveforms determined by the dispersion and pump spectrum.
This formalism also enables us to calculate and engineer the Gaussian entanglement structure of broadband squeezed states of light generated in such processes~\cite{Quesada2022, Patera2010, Brecht2015, Brecht2011, Eckstein2011, Serino2023}.

On the other hand, for \emph{saturated} dynamics, even classical reduced models for the design and control of dispersion relations for useful behaviors are a relatively recent topic of study.
Recent experimental work has rapidly progressed towards a scale where an interplay between saturation and quantum dynamics becomes more relevant.
In the context of traveling-wave devices with $\chi^{(2)}$ nonlinearities, early efforts have focused on enhancing the interaction lengths for pulsed SHG by utilizing dispersion engineering to group-velocity-match the fundamental and second harmonic; experimental examples include SHG in both silicon~\cite{singh2020broadband} and silicon nitride~\cite{hickstein2019self} with field-induced nonlinearities.
More recent work in TFLN managed to eliminate multiple dispersion orders at once to realize quasi-static nonlinear interactions, where each \emph{time-slice} of the pulse acts as an independent temporal mode. Subsequent efforts have demonstrated quasi-static OPA~\cite{ledezma2022intense}, OPG~\cite{Jankowski2022}, and OPOs with quasi-static gain sections~\cite{sekine2023multioctave}.
These devices achieve saturated behaviors with femtojoules, rather than the picojoule scale typical of ultrafast devices. 
As these devices continue to scale to lower photon number, multimode Gaussian models can be self-consistently extended to capture some nonlinear effects induced by saturation~\cite{Ng2023}, as suggested by the hierarchy of scales discussed in Sec.~\ref{sec:hierarchy-mesoscale}.
Such modeling techniques can help guide efforts to engineer quantum noise and Gaussian entanglement in nonlinear ultrafast devices, at least up to the cusp of the classical-to-quantum transition.

While quasi-static devices can achieve saturation with record-low energy, Ref.~\cite{Yanagimoto2022_temporal} points out that these devices operate in a highly multimode limit.
From a quantum perspective, this multimode behavior suggests that quasi-static devices might in fact decohere very rapidly, thus underscoring an important distinction between ``classical'' and ``quantum'' considerations for dispersion engineering.
In the latter case, we must pay close attention to how dispersion \emph{and} nonlinear saturation affect the way photons scatter amongst different ``pulsed supermodes''.
These interactions directly affect how quantum states of broadband pulses couple to undesired modes and effectively decohere.
Proposals for mitigating these effects, such as the temporal trapping~\cite{Yanagimoto2022_temporal} scheme discussed in Sec.~\ref{sec:quantum-multimode}, rely on complicated multi-wave interactions that simultaneously leverage cross-phase modulation, large group-velocity dispersion, and group-velocity matching between several co-propagating pulses in order to confine a ultrashort-pulse quantum state to a single mode.
In this context, it becomes clear that new design strategies are needed to realize control over multiple dispersion orders at many disparate wavelengths.

To this end, engineering the cross-section of a simple ridge waveguide alone is not sufficient.
Instead, we expect next-generation devices to leverage any of a number of emerging new techniques for dispersion engineering.
These include inverse-dispersion engineering of photonic crystal waveguides~\cite{Vercruysse2020}, corrugated waveguides~\cite{Lucas2023}, and coupled-core waveguides~\cite{Guo2020}.
At heart, all of these techniques share the same underlying principle; coupling between two independent propagating modes modifies the dispersion relations of the emergent supermodes.
In the case of Refs.~\cite{Vercruysse2020, Lucas2023}, a forward-propagating mode can be coupled to a backward propagating mode either by engineering the pattern of holes cut into a photonic crystal, or by engineering the Fourier components of a weak surface corrugation of the waveguide.
In the case of~\cite{Guo2020}, dispersion engineering occurs by hybridizing the forward-propagating modes of two waveguides.
At the time of this writing, these approaches have not yet been employed to control the dynamics of $\chi^{(2)}$ devices, and this subject is a wide-open frontier.

\subsection{Building out our quantum-control toolbox}
\label{sec:quantum-toolbox}

The need to manipulate, detect, and control ultrafast mesoscopic optical signals presents a unique combination of challenges pointing towards new experimental capabilities in quantum optics.
By nature, the quantum states and signals in this regime simultaneously possess important classical components that must be precisely controlled alongside fragile quantum features that behave nonintuitively under measurement and manipulation.
On top of this, the manifestation of such states in the form of ultrafast pulses means that these classical and quantum features are encoded, variously, in femtosecond-scale waveforms, terahertz-scale bandwidths, and thousands of optical channels.
The control of classical dynamics, manipulation of quantum states, and high-capacity processing of optical signals are all experimentally mature capabilities in the fields of nonlinear optics, quantum optics, and optical communications, respectively.
The challenge lies in unifying these existing competencies to generate, manipulate, and process mesoscopic quantum features.

\paragraph{Isolating and observing non-Gaussian features}
Compared to classical devices, the ``measurement chain'' for quantum devices can be much more involved, as quantum features occupy a higher-dimensional space often only accessible via phase-sensitive, mode-selective, and high-efficiency detection methods, such as Wigner tomography with a mode-matched, phase-stabilized local oscillator~\cite{Lvovsky2001} or with photon-number-resolving detectors~\cite{Banaszek1996,nehra2019state,laiho2009direct}.
But while such measurements are routinely done in cavity-QED optical systems, mesoscopic quantum states also possess a strong classical component, which, as discussed in Sec.~\ref{sec:Gaussian-Hamiltonian-engineering}, often plays a key role in the functionality of the device.
Because non-Gaussian features are expected to manifest alongside and in the context of Gaussian ones, direct access to non-Gaussianity likely requires the experimental capability to ``cancel out'' classical features, e.g., using coherent- or squeezed-state (or perhaps even nonlinear~\cite{Yurke1986}) interferometers; such design patterns need to separate non-Gaussian features from classical ones at the output while still allowing dynamical interplay between them in the intermediary.
Finally, it is also worth mentioning that, more generally, the detection, characterization, and validation of non-Gaussian states, e.g., the calculation and measurement of non-Gaussian witnesses, constitute an ongoing field of research, particularly in the ultrafast and multimode-entangled setting~\cite{Jayachandran2023,Hughes2014,Chabaud2021,Walschaers2021}.

\paragraph{Optical multiplexing of multimode quantum states}

As discussed in Sec.~\ref{sec:multimode}, one of the most exciting but challenging aspects of working with optical signals, either quantum or classical, is the massive number of modes on which those signals can be carried.
For example, optical degrees of freedom may exist in the time domain, distinguished by arrival times, or frequency domain, distinguished by wavelength, or any of a multitude of other encodings.
Classically, multiplexing optical signals into and out of such encodings is a straightforward application of signal processing.
In the quantum regime, however, these degrees of freedom can (and often are) strongly entangled, and the process of (de)multiplexing involves quantum interference that reshuffles the entanglement structure of the state relative to the encoding.
Thus, it may be necessary to demultiplex quantum signals \emph{all-optically}, i.e., for the signal processing to happen \emph{before} measurements.
Partially addressing this need, recent years have seen significant advancements in the field of ``quantum pulse gating''~\cite{Brecht2015, Brecht2011, Eckstein2011, Serino2023}, where photonic qubits are jointly encoded in temporal modes multiplexed via mode-selective optical techniques based on optical frequency conversion.
As in all experimental capabilities, the challenge will be in scaling up these techniques, especially in the context of fully on-chip architectures that demand small footprint.
Frequency-domain techniques require the use of complex and bulky dispersive elements, such as pump shapers, wavelength-division-multiplexers, and reconfigurable optics.
On the other hand, optical multiplexing in the time domain often requires long delay lines (for $N$ time bins, at least $N$ delays with lengths scaling as $N$).
Furthermore, any losses or noise incurred in complex all-optical multiplexing operations directly translate into irrecoverable decoherence of the fragile non-Gaussian features in the quantum state.

\paragraph{Coherent manipulation of mesoscopic optical signals}

The hybrid quantum-classical nature of mesoscopic states suggests that to realize their full potential, we likely require new ways of manipulating them that are neither purely classical nor quantum.
For example, we might want to \emph{encode} information in a classical fashion, e.g., as the time-varying value of some simple observable, but we might want to \emph{process} that information in a way that preserves quantum properties, i.e., using quantum channel operations that preserve entanglement and interference.
Furthermore, all-optical manipulation inherently preserves the full bandwidth of optical signals and allows THz-scale operation without electronic bottlenecks.
Recent years have seen rapid development of passive linear functions---beamsplitter networks, fan-in/out, phase-stable links, etc.---though technical scaling challenges remain, and a concerted effort is needed to develop these components on material platforms compatible with nonlinear photonics. 
On top of this, the manipulation of mesoscopic signals will likely also require \emph{nonlinear} optical processing; these functions can be classical, such as coherent switches and analog multipliers, or nonclassical, such as quantum nondemolition readout.
Moving forward, we expect to see a rapid proliferation of nanophotonic devices that leverage efficient nonlinear optics to address these needs.
To wit, perusing a catalog of plausible additions to our engineering toolbox:\vspace{-5pt}
\begin{itemize} \addtolength\itemsep{-7pt}
    \item \textit{Ultrafast optical switching and gating}: Ref.~\cite{Guo2022} showed that dispersion-engineered waveguides can be nonlinearly coupled, with potential for fJ-level operation and beyond.
    More sophisticated switches, perhaps leveraging techniques like adiabatic frequency conversion~\cite{Inoue2023}, may eventually enable all-optical routing and timing of signals on chip, for example, to implement the gated resonators mentioned in Sec.~\ref{sec:experiment-tradeoffs}.
    \item \textit{Electronics-free detection and feedback}: Degenerate OPAs can act as all-optical and high-bandwidth probes to access phase-sensitive information without the use of balanced homodyne detection~\cite{Shaked2018, Takanashi2020, Nehra2022few, Kalash2023}.
    The re-encoding of sensitive quantum signals into the antisqueezed quadrature also makes these schemes remarkably robust against loss (see Sec.~\ref{sec:photonic-qip} for potential applications).
    More generally, new design patterns are needed to cascade and feed back signals from all-optical detection schemes like non-Gaussian quantum nondemolition measurements (see Sec.~\ref{sec:qnd}).
    \item \textit{Optical units of computation}: Beyond the holy grail of the digital (bistable) optical transistor, computation with mesoscopic nonlinear optics will likely incorporate features of analog and CV quantum information processing as well, e.g., to support computational paradigms for photonic machine learning~\cite{Li2022}; see also Sec.~\ref{sec:top-down}.
    Cascading efficient sum/difference-frequency generation can be one way to realize all-optical multiply-accumulate operations~\cite{Reifenstein2023}.
    Ultra-low-threshold optical parametric amplifiers and oscillators operating in the mesoscopic regime naturally provide a source of quantum nonlinearity~\cite{onodera2022nonlinear} to generate and manipulate mesoscopic signals.
\end{itemize}

\paragraph{Nonlinear dynamics as a design framework}

Finally, it is worth emphasizing that, at its very core, ultrafast nonlinear optics is rooted in \emph{dynamics}, and, as described further in Sec.~\ref{sec:augumented}, many useful functionalities in mesoscopic nonlinear-optical devices will likely draw upon \emph{emergent} behavior in quantized coupled wave equations.
To enable quantum engineering in this setting, we must develop comprehensive device models and design principles for determining how multimode dynamics affect the development of non-Gaussian physics.
We have already seen that dispersion can affect the entanglement structure in multimode squeezing, that the Gaussian dynamics of these squeezing supermodes can affect (and potentially enhance) the formation of non-Gaussian features, and that nonlinear effects can lead to effective decoherence, among many other examples.
On the hardware level, these considerations propagate directly back to the \emph{design} of each device, e.g., into questions about waveguide design for dispersion engineering, nanopatterning of electrodes for periodic poling, design of couplers to properly route signals, and so on.
On the modeling side, coupled wave equations are not limited to simply describing multimode optical parametric interactions but can also include dynamics arising from Raman scattering, electro-optic effects, optical cascades, interactions with atomic systems, etc.
These effects are often taken for granted in classical design, but subtleties in their quantum modeling certainly exist (see Ref.~\cite{Gustin2023} for a recent example), so how should we go about reducing the correct quantum models to a point where they are tractable enough to fit into design workflow?
Ultimately, a successful framework for mesoscale nonlinear-optical engineering needs to draw upon sophisticated, multi-scale, and dynamical models of both classical and quantum device physics and integrate such tools to span the entire device-design cycle, from mechanical CAD, to electromagnetic simulation, to classical-quantum dynamics, and finally to specification and function.

\section{New opportunities at the mesoscale}
Having oriented ourselves to the physics of mesoscopic quantum nonlinear optics and considering some of the conceptual and experimental challenges involved, in this section, we take the liberty of speculating on some new opportunities ahead, revisiting established paradigms in quantum optics to see what the mesoscale has to offer.

\subsection{Quantum nondemolition measurements}

\label{sec:qnd}
As hinted in Sec.~\ref{sec:quantum-toolbox}, nonlinear optics can offer unique opportunities for quantum measurement with strong enough nonlinearities to access the mesoscale.
In particular, nonlinear-optical interactions can realize so-called quantum nondemolition (QND)~\cite{Imoto1985,Grangier1998} measurements, known to provide a powerful means for processing and manipulating quantum information.
At a high level, information about an signal observable is coherently encoded onto a probe light, and the backaction induced by a subsequent probe measurement is minimal in a particular sense~\cite{Grangier1998}.
Furthermore, by reusing the same probe across a sequence of QND measurements, one can, for example, establish entanglement among distant nodes upon measurement~\cite{Nemoto2004}.

Traditionally, due to weak optical nonlinearities, only QND measurements of linear observables (i.e., quadrature amplitudes) have been accessible~\cite{Yoshikawa2008}.
However, the advent of mesoscopic nonlinear optics may enable QND measurements of \emph{nonlinear} observables (e.g., photon number), significantly extending the space of QND functionality.
In particular, by tailoring the interplay between mean-field, Gaussian, and non-Gaussian dynamics in this regime, novel types of QND observables can be shown to arise~\cite{Yanagimoto2023-qnd,Yanagimoto2023-cubic} (see also Box~\ref{box:example-mesoscopic-opa} for more detail). We can even realize measurements that have no known electronics-based realizations (see Box.~\ref{box:xppx}). Multimode physics could further enlarge the space of possibilities.
Reference~\cite{Yang2020} proposes a trapped-ion quantum simulator to implement the QND measurement of a many-body Hamiltonian $\hat{H}$, in which measurement of the probe projects the system to an energy eigenstate of $\hat{H}$.
When applied to optics, QND measurement of such many-body observables could find applications in multimode quantum state engineering, sensing, and sampling~\cite{Novo2021}. 

We also emphasize that the concept of ``nonlinear-optical quantum measurement'' likely encompasses more than just QND measurements: More generally, nonlinear-optical operations can be thought of as providing a form of pre-measurement processing, coherently revealing information embedded in the state that might be challenging to obtain otherwise~\cite{Epstein2021}.
A classical example is frequency-resolved optical gating~\cite{Trebino1997}, where frequency conversion extracts ultrahigh-resolution information about the waveform of an ultrashort pulse and encodes it into the intensity of the output field, which can then be read out by a slow detector.
In electro-optic sampling, the electric field profile of difficult-to-measure (e.g., terahertz) waves is encoded into easy-to-measure optical fields via electro-optic interactions~\cite{Wu1995}.
We anticipate that general mesosopic nonlinear interactions can perform analogous forms of coherent information processing in a quantum-augumented feature space.

\subsection{CV nonlinear-optical quantum computation}
\label{sec:photonic-qip}

In linear-optical quantum computation (LOQC)~\cite{Knill2001, Obrien2007, Arrazola2021, Zhong2020, Takeda2019}, the mainstream approach to optical quantum computation, the non-Gaussian resources necessary for universal quantum gates~\cite{Bartlett2002-GK} are realized by effective nonlinearities induced by photon-counting measurements.
However, the inherently probabilistic nature of such gate operations, the slow speed of single-photon detectors, and the need for cryogenics critically limit the scalability of existing architectures.
As an alternative, nonlinear-optical quantum computation (NLOQC) aims to realize deterministic and measurement-free non-Gaussian quantum operations by directly using nonlinear optics.
While NLOQC is traditionally formulated for single-photon qubits undergoing microscopic nonlinear-optical dynamics~\cite{Chuang1995, Nemoto2004, Milburn1989,Langford2011}, recent studies have shown \emph{mesoscopic} nonlinear optics can offer unique opportunities for NLOQC~\cite{Yanagimoto2020, Yanagimoto2023-qnd}, specifically in the context of CV quantum computation~\cite{Braunstein2005,Takeda2019}, which is the most natural formalism for manipulating quantum information in mesosopic systems. 

The non-Gaussian quantum operations enabled by mesoscopic nonlinear optics can be weaved into existing architectures, e.g., cluster-state-based quantum computation~\cite{Menicucci2006}, using quantum teleportation~\cite{Furusawa1998} and feedforward operations~\cite{Sakaguchi2023}. Notably, mesoscopic nonlinear optics may offer a means to achieving the long-standing goal of all-optical quantum computation, enabling nonlinear-optical quantum measurements (see Sec.~\ref{sec:qnd}), all-optical quantum teleportation~\cite{Ralph1999}, and optical computation~\cite{hughes2019wave} (also see Sec.~\ref{sec:top-down}). Such mechanisms can be made free from electronics, potentially enabling full-scale quantum operations at terahertz bandwidths.

Here, the main challenges lie in properly utilizing the double-edged sword of multimode physics inherent to nonlinear optics.
On one hand, multimode interactions act as effective decoherence channels, hindering high-fidelity quantum gate operations in na\"ive nonlinear-optical implementations~\cite{Shapiro2006}, and elaborate techniques are required to realize high-fidelity gates~\cite{ Yanagimoto2022_temporal,Xia2016, Viswanathan2018, Babushkin2022}.
On the other hand, the large bandwidth of optical fields enables highly multiplexed information encodings, e.g., in frequency~\cite{Roslund2014}, spatial~\cite{Devaux2020}, or temporal~\cite{Asavarant2019, Takeda2017} degrees of freedom.
As we develop more techniques to manipulate multimode quantum states of light~\cite{Brecht2015, Brecht2011, Eckstein2011, Serino2023} and more intuition for mesoscopic quantum dynamics, it is not inconceivable we may discover a massively scalable approach to non-Gaussian quantum gates, uniquely enabled by multimode nonlinear optics.

\onecolumngrid
\vspace{0.2cm}
\begin{aside}[Mesoscopic OPA for QND measurement of $\hat{x}\hat{p}+\hat{p}\hat{x}$] \label{box:xppx}

In Ref.~\cite{Oh2019}, it is shown that a non-Gaussian measurement in the basis of operator $\hat{x}\hat{p}+\hat{p}\hat{x}$ is useful for optimally estimating the phase shift of a Gaussian state.
However, a concrete means of realizing such a measurement has not yet been proposed to our knowledge.
Here, we show how a QND measurement of $\hat{x}\hat{p}+\hat{p}\hat{x}$ can be realized by pumping a mesoscale OPA with a \emph{squeezed state}.

\begin{wrapfigure}{r}{0.5\textwidth}
    \begin{center}
        \vspace{-0.8cm}
\includegraphics[width=0.5\textwidth]{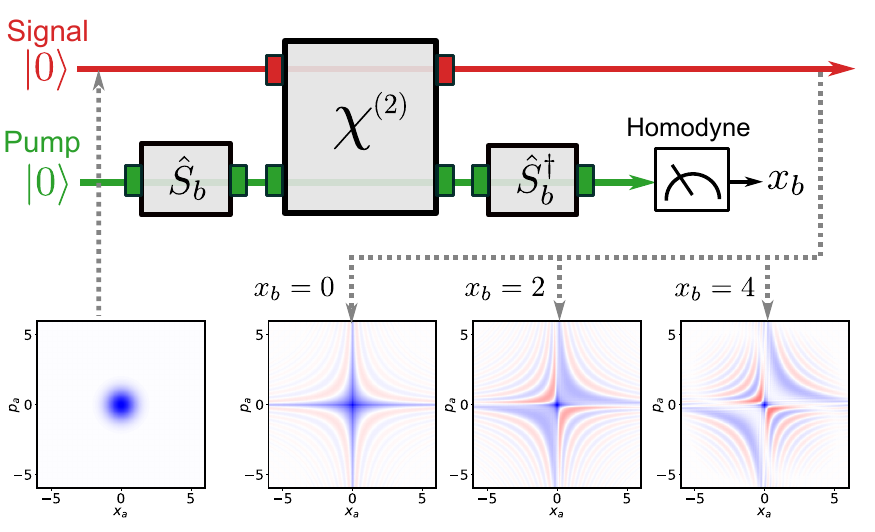}
\vspace{-0.8cm}
\end{center}
  \caption{Schematic of a QND measurement of $\hat{x}\hat{p}+\hat{p}\hat{x}$ using mesoscopic OPA and Gaussian Hamiltonian engineering. Conditional on homodyne measurement of the pump quadrature $\hat{x}_b$, the signal state is projected to various post-measurement states that have features related to the eigenstates of $\hat{x}\hat{p}+\hat{p}\hat{x}$~\cite{Bollini1993}. Bottom images show Wigner functions of the pre-measurement state (leftmost) and post-measurement states. Here, $\lambda=10$ and $gt=0.4$.} \label{fig:xppx}
\end{wrapfigure}

The construction is shown in Fig.~\ref{fig:xppx}.
We consider again the phase-matched OPA interaction $\hat{H}=\frac{1}{2}g(\hat{a}^{\dagger2}\hat{b}+\hat{a}^2\hat{b}^\dagger)$, acting for a time $t$.
Before and after the OPA interaction, we apply a pair of opposite squeezing operations $\hat{S}_b$ and $\hat{S}_b^\dagger$, respectively.
These squeezing operations are defined by their action $\hat{S}_b^\dagger\hat{b}\hat{S}_b=\lambda^{-1}\hat{x}_b+\mathrm{i}\lambda\hat{p}_b$, where $\lambda>1$ is the squeezing factor.
The total unitary of the system is
\begin{align}
    \hat{S}^\dagger_be^{-\mathrm{i}\hat{H}t}\hat{S}_b=e^{-\mathrm{i}\hat{H}_\mathrm{eff}t},
\end{align}
for an effective Hamiltonian
\begin{align}
\hat{H}_\mathrm{eff}=g\lambda(\hat{x}_a\hat{p}_a+\hat{p}_a\hat{x}_a)\hat{p}_b+\mathcal{O}(\lambda^{-1}).
\end{align}
In the limit of strong squeezing $\lambda\gg 1$, the dominant dynamics under $\hat{H}_\mathrm{eff}$ is to induce $x$-displacement of the pump mode, i.e., $\hat{x}_b(t)=\hat{x}_b(t=0)+$ $\frac{g\lambda}{2}(\hat{x}_a\hat{p}_a+\hat{p}_a\hat{x}_a)$. Thus, by measuring the $x$-displacement of the pump, we can infer the value of $\hat{x}_a\hat{p}_a+\hat{p}_a\hat{x}_a$, realizing a QND measurement.

In the simplest construction where the initial pump state is vacuum (as depicted in Fig.~\ref{fig:xppx}), the system can be seen as an OPA pumped by a squeezed-vacuum state.

\end{aside}

\twocolumngrid

\subsection{Quantum-augmented complex wave phenomena} \label{sec:augumented}

Wave equations are the physical basis for some of the most complex dynamical phenomena in nature.
As partial differential equations (PDEs), Maxwell's equations in nonlinear media span wide dynamic ranges in time, frequency, and space, allowing them to naturally manifest multi-scale and emergent physics.
Such spatiotemporal phenomena, from a veritable zoo of solitons~\cite{Grelu2012, Kivshar2003} to the emergence of thermodynamic features~\cite{Wu2019} to literal chaos~\cite{Sciamanna2015}, makes classical nonlinear optics a clean and attractive experimental platform to study the physics of complex systems.
The mesoscopic regime has interesting implications for this PDE-driven view of nonlinear optics.
In the context of Sec.~\ref{sec:mesoscopic}, the classical PDEs responsible for all the complexity discussed above constitute the dynamics of \emph{only} the first-order moments of the quantum state, meaning that the remaining (quantum) observables comprise entirely new dynamical variables that can \emph{augment} the complexity of the classical dynamics.

This can be seen most readily in quantum models based on truncated cumulant expansions (as discussed in Sec.~\ref{sec:model-reduction}), where the higher-order moments of the quantum state are viewed as dynamical quantities alongside the mean field~\cite{Huang2022, Schack1990, Huang2023}.
For example, Ref.~\cite{Ng2023} shows that one can self-consistently derive coupled-wave equations for all the Gaussian (first- and second-order) moments of the quantum state of a pulse as it propagates in a nonlinear medium.
These Gaussian wave equations can be seen as straightforward generalization of the classical ones: Starting with a classical system of $M$ interacting mode amplitudes, we can arrive at a new system of $O(M^2)$ interacting moments.
Importantly, however, they also take a step beyond the linearized approximations usually taken in Gaussian quantum optics, as the dynamics of the second-order moments are \emph{nonlinearly coupled} to the mean field, meaning such models can, in principle, exhibit complex behavior in a quadratically larger ``configuration space'' than classically possible.
As the contributions from quantum fluctuations become more significant, this configuration space enlarges rapidly; there are $O(M^n)$ interacting moments when up to $n$th-order moments contribute.
Physically, the evolution of such higher-order moments describes the dynamics of quantum noise and its correlations, including entanglement, in the state of the lightwave.
Finally, the presence of mesoscopic non-Gaussian features~\cite{Gao2023} (e.g., Wigner negativity) would further augment the complexity accessible by the dynamics, adding elements of e.g., quantum contextuality~\cite{Booth2022,Spekkens2008} and wavefunction interference~\cite{Hong1987}, that no classical theory cannot explain.

It is interesting to consider to what extent such augmented dynamics can lead to distinctively new phenomena in nonlinear optics.
For instance, optical solitons, which are classically stable waveforms, are known to become unstable and diffuse (or evaporate) due to quantum fluctuations~\cite{Bao2021, Villari2018, Seibold2022}, a phenomenon whose explanation has required semiclassical theories of solitons.
Conversely, analyzing the quantum-augumented equations of motion for soliton propagation may then point towards novel types of solitons (not based on coherent or squeezed states, for example) that are more robust to quantum diffusion.
Another possible direction is to study systems where classical dynamics exhibit high sensitivity to initial conditions; under such conditions, it may be possible for even mesoscopic (or even microscopic) quantum features to have an outsized effect on macroscopic dynamics.
For instance, it has been demonstrated that the bifurcation dynamics of an OPO can be sensitively biased by even single-photon-level quantum fluctuations~\cite{Roques-Carmes2023}.
In the most extreme setting, chaotic dynamics exhibit exponential sensitivity to initial conditions, but it is also well known classically that intermediate between stable and chaotic regimes, there can exist sweet spots at the ``edge'' of chaos~\cite{Langton1990}. Here, the dynamics are sensitive enough that small quantum perturbations can give rise to rich and emergent phenomena, yet remain insensitive enough that they can still be controllably engineered.

\subsection{Non-Gaussian quantum light-matter interactions}
\label{sec:light-matter}
The advent of coherent light generated by lasers revolutionized our ability to study the interactions between light and matter, triggering rapid developments in both basic science and technology.
For instance, coherent light is able to efficiently drive narrowband optical transitions in atoms and molecules, enabling core aspects of modern atomic, molecular, and optical (AMO) physics.
As coherent states comprise just one specific subclass among many different quantum states of light, what \emph{more} could we do with access to an even greater variety of states?
In this context, there is growing interest in how the quantum nature of light changes the ways optical and matter degrees of freedom interact, with examples of pioneering work in atom-cavity QED~\cite{Leroux2018,Qin2018}, molecular spectroscopy~\cite{Michael2019}, second-harmonic generation~\cite{Walls1972}, and ultrafast electron dynamics~\cite{Tzur2023}.

For example, Ref.~\cite{Gorlach2023} studied high-harmonic generation driven by coherent, squeezed, Fock, and thermal states, finding that the threshold critically depends on the photon statistics of the pump. Such quantum phenomena can lead to novel functions. When an OPA is driven by a squeezed state (instead of a coherent state), we can realize a QND measurement of a product operator $\hat{x}\hat{p}+\hat{p}\hat{x}$, which is completely different from the conventional function of parametric amplifier (see Box.~\ref{box:xppx}).
At the same time, it is worth noting that the states mentioned here constitute only the more well known (and named) classes of quantum-optical states, while there clearly exist many (perhaps uncountably many) more ``unnamed'' non-Gaussian states, each of which potentially interacts with matter in a unique way.

Moreover, the volume of this state space increases rapidly with the number of modes, suggesting that multimode physics, in particular, plays an important role in opening up this space.
For example, specific spatiotemporal entanglement structures in of quantum light might act as efficient probes into non-local structure in quantum matter.
This can be seen concretely in superabsorption, where an atomic cloud selectively absorbs light with a particular temporal entanglement structure (i.e., a superradiant photonic state) much more strongly than other states of light~\cite{Yang2021}.
The advent of new quantum light sources based on mesoscopic ultrafast optics could enable a new frontier for studying quantum light-matter interactions spanning over the entire electromagnetic spectrum.

\subsection{Nonlinear dissipation in multimode photonics}
Generally, linear dissipation (i.e., photon loss), which decoheres quantum states, is viewed as detrimental to quantum experiments.
On the other hand, \emph{nonlinear} dissipation can generate non-classical features and be a valuable resource for quantum engineering, even enabling universal quantum computation without any unitary components~\cite{Verstraete2009}.
In nonlinear photonics, Ref.~\cite{Rivera2023} shows that nonlinear dissipation of an optical cavity can be engineered so that the cavity is only loss-free when a specific number of photons are present; consequently, the system deterministically evolves into a Fock state.
In a singly-resonant degenerate OPO, two-photon loss induced by pump depletion above threshold has long been known to generate Schr\"odinger cat states via dissipative dynamics~\cite{Wolinsky1988}.

More generally, in the theory of open quantum systems, dissipation arises whenever we partition our physics into ``system'' and ``reservoir'' degrees of freedom, followed by an assumption that we have no access to (i.e., we partial-trace out) the reservoir.
While the system-reservoir boundary is often naturally identified, e.g., a mirror separating an optical cavity and the rest of the world, the idea is more general~\cite{Mabuchi2012}:
In a multimode quantum field, we can just as well identify certain modes (e.g., a particular waveform or band of frequencies) as the ``system'' with all the other ``irrelevant'' modes as the reservoir, and nonlinear coupling between the two can therefore lead to notions of nonlinear dissipation.
For instance, for broadband $\chi^{(2)}$ interactions in a waveguide, second-harmonic (SH) modes are nonlinearly coupled to a continuum of fundamental-harmonic (FH) modes. Thus, the parametric downconversion of SH photons to FH can, in certain circumstances, be viewed as dissipative decay to a continuum reservoir of FH modes~\cite{Gustin2023}.
Other effects in quantum ultrafast optics, such as quantum soliton evaporation, may also be understood as a manifestation of effective system-reservoir physics in this sense.
Notably, depending on the nature of the multimode nonlinear couplings, such system-reservoir interactions can even be non-Markovian~\cite{Sloan2023}, giving rise to rich dynamical features which may be difficult to engineer otherwise.
Because mesoscopic ultrafast nonlinear optics features a unique combination of strong nonlinearity, inherent multimodedness, and Gaussian engineerability of the interactions, we expect it to serve as an interesting platform to explore novel forms of nonlinear dissipation and ideas for how to engineer them.

\subsection{A top-down approach to optical computing at the mesoscale}
\label{sec:top-down}
Despite decades of research, photonics has yet to replace or even complement the singular role that CMOS electronics occupy in digital computation.
The crux of the challenge is there are several and serious engineering obstacles to realizing a competitive \emph{optical transistor}, at least one that posseses all the desired properties of the CMOS incumbent~\cite{Miller2010}; this holds true even considering the rapid improvements in energy scale reflected in Fig.~\ref{fig:devices}.
In fact, there is little reason to believe that the transistor is the most natural \emph{abstraction} of how a photon ``computes''~\cite{Hooker2020}; comparing an abstract photonic device to its electronic counterpart, they are quite disparate, from their typical length scale (microns vs.\ nm) and time scale (THz vs.\ GHz) to the nature of their interactions (linear vs.\ Coulomb).
Rather than trying to enforce the bottom-up approach to computing borrowed from semiconductors---transistors, instruction sets, and algorithms---we should instead reconsider which computational paradigms might be more natural for light.

In this context, the recent revolution of machine learning and neuromorphic computing as viable alternatives to traditional computation represents an opportunity to revisit optical computation from a \emph{top-down approach}.
That is, high-level and hardware-agnostic techniques such as training~\cite{wright2022deep} and feedback~\cite{Shimazu2021, Yamamoto2014} can be used to fulfill specific and powerful \emph{tasks}, in contrast to compiling programs to implement \emph{algorithms}.
(It is also worth noting that the tasks themselves are quite general, encompassing both classical and quantum objectives; see, for example, some recent work in top-down engineering of quantum gates~\cite{Eickbusch2022, Krastanov2021}.)
In terms of using physical systems for top-down computing, one rudimentary example can be seen in the work surrounding coherent Ising machines~\cite{mcmahon2016fully, inagaki2016coherent, leleu2017combinatorial}, where a feedback loop harnesses the optical dynamics of a nonlinear oscillator network to approximately solve hard combinatorial optimization problems, using concurrent applications of linear coupling and local nonlinearities in a fashion reminiscent of recurrent neural networks and neural ODEs~\cite{Chen2018}.
Another example, explicitly evoking the concept of training, can be seen in recent attempts to use the complex wave dynamics of nonlinear waveguides as a reservoir computer~\cite{marcucci2020theory}.
Such systems can potentially achieve unprecedented levels of computational throughput by combining the native speed of all-optical reservoirs with the prospect of directly training parameters of the physical system (e.g., the refractive index distribution, via electro-optic effects) to eliminate digital pre- or post-processing~\cite{hughes2019wave, papp2021nanoscale, wright2022deep}.

We argue that the same top-down approach can be used to harness \emph{quantum} behavior for computation as well, at least in the mesoscopic regime.
As discussed in Sec.~\ref{sec:augumented}, quantum features such as non-Gaussianity can be seen as simply increasing the configuration space of a physical system, in a sense placing quantum and classical features on equal footing as resources.
In contrast to traditional quantum computing, where the realization of a desired quantum algorithm depends critically on the \emph{precise} structure of a quantum circuit, the use of feedback or training in top-down computing can make them inherently robust to variations in the underlying physics, even when complex multimode and mesoscopic dynamics are involved.
In this way, although we likely give up the theoretical possibility of obtaining provable ``exponential speedups'', we gain a straightforward way to deploy quantum mechanics for computation; when properly trained or controlled, such ``quantum-augmented machines'' almost self-evidently have the potential to improve the performance of learning tasks (in terms of time or energy consumed). 
Their physical embodiment as a quantum-optical system also enables coherent interactions with the room-temperature environment, expanding the scope of signals they can process.

\subsection{Quantum sensing and metrology}
By leveraging the quantum nature of photons in sensing and metrology, one can achieve performance that exceeds classical limitations.
Broadly speaking, such quantum advantage can arise from the use of either (1) non-classical states, or (2) non-classical measurements.
A well known example of (1) is the Laser Interferometer Gravitational-Wave Observatory (LIGO), where using a squeezed state of light enabled them to push the measurement noise below the classical limit~\cite{LIGO2013}.
Depending on the parameter one wishes to measure, certain quantum states are known to provide quantum advantage, e.g., squeezed states for phase shifts and the grid states for general displacements~\cite{Duivenvoorden2017}.
Interestingly, quantum advantage can also be realized using (2), non-classical measurements, even when the probe state is completely classical.
This is exemplified in Ref.~\cite{Tsang2016}, where it is shown that the separation between two incoherent light sources can be measured beyond the classical resolution limit using spatially multiplexed photon-counting measurements.
In this approach, it is essential to perform non-Gaussian quantum measurements in a specific spatial or temporal mode basis.

Mesoscopic nonlinear optics provide unique ways of realizing both types of resources for quantum sensing and metrology.
In quantum light sources, mesoscopic devices can not only deterministically generate well known non-Gaussian quantum states but also ``unnamed'' non-Gaussian states as well, which could provide new quantum advantages for precision spectroscopy in light-matter interactions (see also Sec.~\ref{sec:light-matter}).
In non-Gaussian measurements, the interplay between Gaussian and non-Gaussian features in mesoscopic nonlinear optics can enable flexible engineering of the measurement basis, generating useful quantum measurements that are challenging to realize otherwise (see Box.~\ref{box:xppx} for a concrete example). Such measurements can be implemented in a QND manner, which is a unique advantage over electronics-based measurements (see Sec.~\ref{sec:qnd}).

\section{Towards quantum optics, in optics}
\begin{figure}[b]
    \centering
    \includegraphics[width=0.5\textwidth]{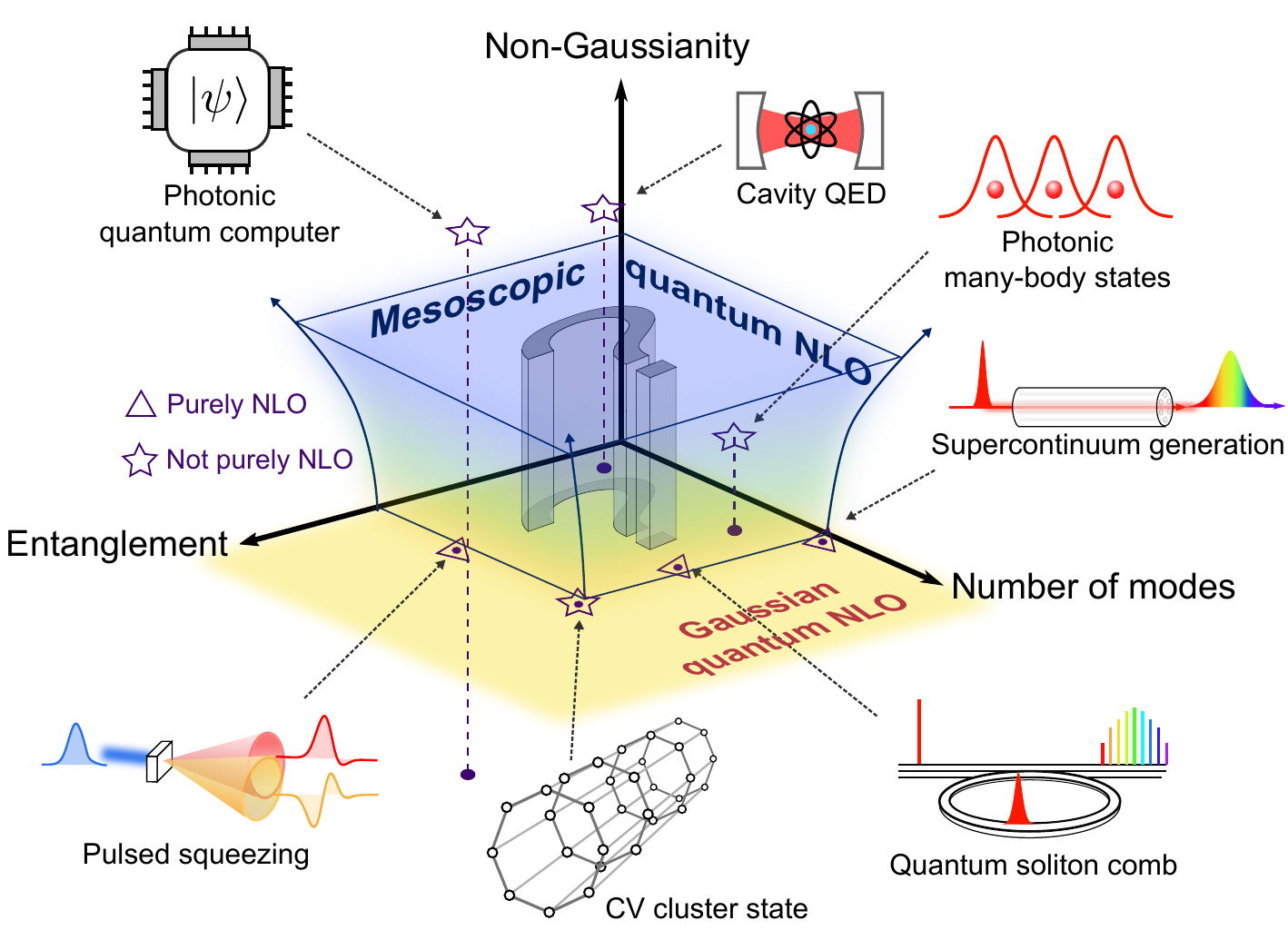}
    \caption{Illustration of physical phenomena and systems with various quantum states of light that they incorporate, where we qualitatively depict the nominal amounts of non-Gaussianity, entanglement, and multimodeness.}
    \label{fig:quantum-states}
\end{figure}
While optical science and technology has always enjoyed a central place in advancing our understanding of quantum mechanics, the field of quantum optics itself has heavily relied on light's interaction with \emph{non-optical} systems (from atoms to superconducting single-photon detectors) for the most critical operations at the heart of quantum engineering.
The progression of nonlinear optics into the mesoscale energy regime, however, offers a chance to disrupt this status quo, potentially opening a route towards non-Gaussian quantum physics with purely nonlinear-optical means, i.e., \emph{quantum optics, in optics}.
As illustrated in Fig.~\ref{fig:quantum-states}, we can view opportunities on this front as extending nonlinear optics into another dimension in the space of possibilities, departing from the limitations of Gaussian quantum states.
Because nonlinear optics has already demonstrated the capability for large bandwidth (or equivalently a large number of modes) and rich Gaussian dynamics (e.g., multimode entanglement), this progression can be expected to naturally engender phenomena at scales of complexity otherwise inaccessible in other non-Gaussian platforms.

While the complex nature of mesoscopic multimode physics may present challenges for constructing a complete engineering framework, it is also our view that this is a prime opportunity to rethink our approach to quantum engineering as a whole; looking at features and dynamics rather than states and gates, or phenomena and functions rather than protocols and algorithms.
Nonlinear optics, historically, has been a field propelled by hardware-driven, often bespoke, solutions to concrete problems, from light sources to metrology and sensing.
We expect this legacy to persist through the classical-quantum transition.

Considering such potential, in this mini-review, we have provided one perspective on how non-Gaussian quantum physics might naturally emerge in the near future of nonlinear optics, and we have introduced theoretical tools for conceptualizing the interplay between Gaussian and non-Gaussian features in mesoscopic quantum dynamics.
We have drawn attention to what we see as important experimental and technical challenges, with implications for advancing our ability to manipulate and exploit quantum ultrafast nonlinear optics in this regime.
Finally, we have also elaborated, with varying degrees of speculation, on some new opportunities to establish new approaches and applications to quantum science driven by mesoscopic, non-Gaussian, and multimode physics, from quantum information and metrology to analog and neuromorphic computing.
While the present status of this effort is admittedly theoretical and speculative, we hope that the topics presented herein can inspire new research and researchers in both theory and experiment at this unique frontier of quantum nonlinear optics.

\section*{Acknowledgements}
The authors wish to thank NTT Research for their financial and technical support. This work is supported by ARO Grant W911NF-23-1-0048, NSF Grants ECCS-1846273, PHY-2011363, and CCF-1918549, AFOSR award FA9550-23-1-0755, and DARPA Grant No. D23AP00158. The authors wish to thank Noah Flemens, Ryoto Sekine, and Akira Furusawa for insightful discussions and feedback. The authors are grateful to Ryoto Sekine for providing the image of their experimental device shown in Fig.~\ref{fig:devices}(c) IV. Custom illustrative figure outset of Fig.~\ref{fig:devices}(b) by Science Journal Editors (www.scienceje.com).

\bibliography{myfile}

\end{document}